\let\mypdfximage\pdfximage
\def\pdfximage{\immediate\mypdfximage}

\documentclass[letterpaper, 11pt]{article}

\usepackage{algorithm}
\usepackage{algorithmic}
\usepackage{amsmath}
\usepackage{amssymb}
\usepackage{amsthm}
\usepackage[font=small]{caption} \usepackage{epsfig}
\usepackage{geometry}
\usepackage{graphicx}
\usepackage[colorlinks=true, citecolor=blue, filecolor=black, linkcolor=red, urlcolor=blue]{hyperref}
\usepackage[caption=false]{subfig}
\usepackage{amsfonts}
\usepackage{epstopdf}
\usepackage{color}
\usepackage[textsize=tiny]{todonotes}

\usepackage{soul} \setstcolor{red}
\setulcolor{blue}

\usepackage[capitalise,nameinlink,noabbrev]{cleveref} 
\renewcommand{\(}{\left(}
\renewcommand{\)}{\right)}

\newcommand{\abs}[1]{\left|#1\right|}

\newcommand{\norm}[2]{\left\| #1 \right\|_{#2}}

\newcommand{\vvvert}{|\kern-1pt|\kern-1pt|}

\newcommand{\bd}{\textbf{d}}

\newcommand{\bu}{\textbf{u}}

\newcommand{\bx}{\textbf{x}}
\newcommand{\by}{\textbf{y}}
\newcommand{\bz}{\textbf{z}}

\newcommand{\bB}{\mathbf{B}}

\newcommand{\bD}{\mathbf{D}}

\newcommand{\bG}{\mathbf{G}}
\newcommand{\bH}{\mathbf{H}}

\newcommand{\bU}{\mathbf{U}}

\newcommand{\bepsilon}{\boldsymbol{\epsilon}}

\newcommand{\btheta}{\boldsymbol{\theta}}

\newcommand{\bxi}{\boldsymbol{\xi}}

\newcommand{\bTheta}{\boldsymbol{\Theta}}

\newcommand{\EE}{\mathbb{E}}
\newcommand{\II}{\mathbb{I}}

\newcommand{\RR}{\mathbb{R}}

\newcommand{\CD}{\mathcal{D}}

\newcommand{\CI}{\mathcal{I}}

\newcommand{\CN}{\mathcal{N}}

\newcommand{\CU}{\mathcal{U}}

\newcommand{\CY}{\mathcal{Y}}
\newcommand{\CZ}{\mathcal{Z}}

\newcommand{\Cov}{\textrm{Cov}}

\newcommand{\argmax}{\operatornamewithlimits{argmax}}

\newcommand{\DKL}{D_{\mathrm{KL}}}

\crefname{section}{Sec.}{Sec.}
\Crefname{section}{Section}{Sections}
\crefname{subsection}{Sec.}{Sec.}
\Crefname{subsection}{Section}{Sections}
\crefname{figure}{Fig.}{Fig.}
\Crefname{figure}{Figure}{Figures}
\crefname{equation}{Eqn.}{Eqn.}
\Crefname{equation}{Equation}{Equations}

\crefdefaultlabelformat{#2\textup{#1}#3}

\floatname{algorithm}{Algorithm}

\title{Goal-Oriented Bayesian Optimal Experimental Design for Nonlinear Models using Markov Chain Monte Carlo}

\author{Shijie Zhong\footnote{\href{mailto:szhong12@jhu.edu}{szhong12@jhu.edu}, Johns Hopkins University, Baltimore, MD 21218, USA.},
    Wanggang Shen\footnote{\href{mailto:wgshen@umich.edu}{wgshen@umich.edu}, University of Michigan, Ann Arbor, MI 48109, USA.}, 
    Tommie Catanach\footnote{\href{mailto:tacatan@sandia.gov}{tacatan@sandia.gov}, Sandia National Laboratories, Livermore, CA 94550, USA.}, 
    Xun Huan\footnote{\href{mailto:xhuan@umich.edu}{xhuan@umich.edu}, University of Michigan, Ann Arbor, MI 48109, USA. \href{https://uq.engin.umich.edu}{https://uq.engin.umich.edu}}}

\date{}

\begin{document}

\maketitle

\begin{abstract}
   Optimal experimental design (OED) provides a systematic approach to quantify and maximize the value of experimental data. Under a Bayesian approach, conventional OED maximizes the expected information gain (EIG) on model parameters. However, we are often interested in not the parameters themselves, but predictive quantities of interest (QoIs) that depend on the parameters in a nonlinear manner. We present a computational framework of predictive goal-oriented OED (GO-OED) suitable for nonlinear observation and prediction models, which seeks the experimental design providing the greatest EIG on the QoIs. 
   In particular, we propose a nested Monte Carlo estimator for the QoI EIG, featuring Markov chain Monte Carlo for posterior sampling and kernel density estimation for evaluating the posterior-predictive density and its Kullback-Leibler divergence from the prior-predictive. The GO-OED design is then found by maximizing the EIG over the design space using Bayesian optimization.
   We demonstrate the effectiveness of the overall nonlinear GO-OED method, and illustrate its differences versus conventional non-GO-OED, through various test problems and an application of sensor placement for source inversion in a convection-diffusion field. 
\end{abstract}

\section{Introduction}
\label{s:introduction}

Experiments play a central role in science and engineering. For example, they can provide data for us to better understand the underlying dynamics of a complex system, to examine process behavior in unexplored regimes, and to validate the performance of a designed engineering system under real-world conditions. 
Different experiments also offer varying degrees of usefulness.
Consideration of an experiment's value thus becomes particularly important when costs are high or resources are limited, such as the case of choosing components %
for destructive testing, deploying sensors to hazardous 
locations, and selecting experiments for flagship scientific facilities and instruments.
The field of optimal experimental design (OED) (see, e.g.,~\cite{Lindley1956,Fedorov1972,Chaloner1995,Atkinson2007,Ryan2016,Alexanderian2021,Rainforth2023,Huan2024}) thus seeks to identify experiments that can provide the greatest value. 

In order to compare and optimize the value of experiments, a prerequisite is to define a utility metric that appropriately and quantitatively reflects the value of an experiment with respect to the experiment goal. 
For example, a common experiment goal is to reduce the uncertainty about unknown model parameters. 
For linear inverse problems~\cite{Atkinson2007}, this uncertainty can be portrayed through the Fisher's information matrix (FIM) that is inversely proportional to the parameter covariance matrix. The well-known alphabetic optimality criteria are then formed via different operations on the FIM (e.g., A-optimality maximizes the trace of FIM, D-optimality maximizes the log-determinant of FIM, etc.). Bayesian versions of the alphabetic optimality criteria can also be formed by inserting the prior covariance~\cite{Chaloner1995}.
For nonlinear models, mutual information (MI)~\cite{Lindley1956} between the parameters and observables (equivalently, the expected Kullback--Leibler (KL) divergence from the Bayesian prior to the posterior) is commonly adopted to measure the expected information gain (EIG) (i.e., expected uncertainty reduction) in the model parameters.
{However, parameter EIG is generally not available in closed form for nonlinear models, and must be estimated numerically. A common approach is the nested MC (NMC) estimator~\cite{Ryan2003}, which uses Monte Carlo (MC) integration to approximate both the expectation and the marginal likelihood in the EIG. To improve the computational efficiency of NMC, techniques based on importance sampling~\cite{Beck2018,Feng2019,Englezou2022} and surrogate modeling~\cite{Huan2013,Duong2023} have also been incorporated. Additionally, other approaches, such as those leveraging Gaussian approximations~\cite{Long2013,Overstall2018} and lower and upper bounds of the EIG~\cite{Barber2003,Poole2019,Foster2019,Kleinegesse2020}, have also been proposed. Another approach is to directly approximate the posterior under different data realizations with 
Markov chain Monte Carlo (MCMC), and then apply linear OED criteria such as A- or D-optimality~\cite{ryan_optimal_2016}.}
In many situations, however, reducing the uncertainty of model parameters 
is not the ultimate goal. Instead, the experiment goal may entail reducing the uncertainty towards a downstream purpose (e.g., estimating the failure probability of a component, predicting the operational envelope of a system, or making a decision that minimizes risk at a future time) that depends on the learned model parameters and their uncertainty. 
Forming and optimizing a criterion reflecting such goal steers OED to be directly relevant to the scientific question which motivated the experimental effort, and can reveal designs {that} significantly differ from OED that does not take these goals into account. 
We refer to such an approach a \emph{goal-oriented} optimal experimental design (GO-OED).  

In this paper, we specifically consider the case where the goal is to reduce uncertainty on \emph{predictive quantities of interest (QoIs)} whose value or distribution can be derived from the model parameters---i.e., a \emph{predictive} GO-OED. Hence, any uncertainty on the model parameters must be propagated to the QoIs through a parameter-to-QoI mapping (i.e., prediction model) that in general differs from the parameter-to-observable mapping (i.e., observation model). Understanding how uncertainty reduction due to experimental data propagates from the model parameters to the QoIs is non-trivial and requires new algorthmic advances.

The simplest forms of alphabetic optimality that involve predictive quantities are the L- and D$_\text{A}$-optimal designs, which respectively optimizes the trace and log-determinant of the covariance under a linear combination of the model parameters~\cite{Atkinson2007}. 
For more general linear prediction models, 
I-optimality minimizes the predictive variance integrated over a region of the prediction model's domain, while V-optimality minimizes over a set of points and G-optimality minimizes the maximum predictive variance over a region~\cite{Atkinson2007}. 
More recent efforts demonstrated gradient-based techniques for tackling Bayesian D$_A$- and L-optimal designs~\cite{Attia2018}, and advanced scalable offline-online decompositions and low-rank approximations for reducing the complexity for high-dimensional QoIs~\cite{wu2021efficient}. These formulations, however, continue to require linearity in the observation and prediction models. 

Nonlinear GO-OED's theoretical formulation originates from~\cite{Bernardo1979}, but computational approaches were not considered. GO-OED's computation is significantly more challenging than non-GO-OED that focuses on the parameter EIG. 
A related effort, the OED for prediction (OED4P) framework~\cite{Butler2020}, has been proposed by introducing a stochastic inverse problem aimed at finding a distribution whose push-forward through the parameter-to-observable mapping matches the observed data distribution~\cite{Butler2018,Butler2018a}. The corresponding update distribution for the parameters is similar to the Bayesian posterior, but replaces the marginal likelihood with the initial push-forward distribution of the predictive QoIs.
As a result of avoiding the marginal likelihood, the update distribution and its push-forward can be easily sampled, and the expected KL divergence from the initial to the update distributions (and for their push-forward counterparts) can be estimated inexpensively. 
Like other data consistent inversion approaches, OED4P is flexible but also struggles to scale with the dimensionality of the data and the number of experiments as the push-forward mapping is harder to approximate.
Moreover, OED4P is built upon principles differing from Bayesian probability, where the latter features posterior distributions that emerge from conditioning on new data instead.
In this paper, we present new computational approaches for a Bayesian predictive GO-OED method that estimates and optimizes the EIG in the predictive QoIs while accommodating nonlinear observation and prediction models. In particular, we propose a 
{new NMC}
estimator for the EIG in the QoIs.
{A key component of this estimator is computing the posterior-predictive probability density function (PDF) of the QoIs. To achieve this, our approach relies on 1) a method to sample the posterior (along with the corresponding posterior-predictive QoIs) and 2) a method to estimate the posterior-predictive PDF from the samples.
We employ 
MCMC~\cite{Andrieu2003,Robert2004,Various2011} for parameter posterior sampling, which is then propagated through the prediction model to generate posterior-predictive samples of the QoIs. We approximate the posterior-predictive PDF using a tuned kernel density estimation (KDE), enabling the computation of the KL divergence from the prior-predictive to posterior-predictive distributions. These choices may be exchanged with other algorithms depending on the problem or computational considerations. 
Notably, while MCMC provides asymptotic convergence to the posterior distribution and is widely studied with numerous variants and implementations, it can become difficult to scale beyond hundreds of dimensions and requires an explicit likelihood. MCMC can be replaced with other posterior sampling algorithms, such as sequential Monte Carlo~\cite{Chopin2020} and approximate Bayesian computation~\cite{Sisson2018}. Similarly, KDE can be substituted with other density estimation methods, such as Gaussian mixture models or non-parametric techniques like $k$-nearest neighbors. 
Finally, the GO-OED design is found by maximizing a MC average of the predictive KL divergence under different samples of potential experimental observations. Since this optimization involves noisy objective, we employ Bayesian optimization (BO)~\cite{movckus1975bayesian,jones1998efficient,pelikan1999boa, shahriari2015taking,frazier2018tutorial,wang2023recent} for its global optimization capabilities, lack of gradient dependence, and suitability for both continuous and discrete design settings.}
{We consider the Bayesian predictive GO-OED formulation for nonlinear observation and nonlinear prediction models based on optimizing the EIG in the predictive QoIs.}
The key novelty and contributions of our work can be summarized as follows.
\begin{itemize}
\item We propose computational methods for approximating the expected utility  for GO-OED through a nested MC estimator powered by MCMC and KDE, and optimizing it with BO.
\item We demonstrate GO-OED on benchmark tests and a convection-diffusion sensor placement application, while contrasting the differences between GO-OED and non-GO-OED results.
\end{itemize}

The remainder of this paper is structured as follows. \Cref{s:formulation} presents the nonlinear Bayesian GO-OED formulation. \Cref{s:methods} details the numerical methods used to solve the GO-OED problem. \Cref{s:results} provides three examples of GO-OED problems to demonstrate our GO-OED method, along with discussions and interpretations of the results. Finally, \cref{s:conclusions} offers concluding remarks, discussions on limitations, and ideas of future work.

\section{Problem Formulation}
\label{s:formulation}

Consider an \emph{observation model} in the form
\begin{align}
    \by = \mathbf{G}(\btheta, \bd) + \bepsilon
    \label{e:obs_model}
\end{align}
where $\by \in \CY \subseteq \RR^{n_y}$ is the observation data, $\btheta \in \bTheta \subseteq \RR^{n_{\theta}}$ the vector of model parameters, $\bd \in \CD \subseteq \RR^{n_d}$ the experimental design vector, $\mathbf{G}: \bTheta \times \CD \to \RR^{n_y}$ 
a nonlinear observation forward model (parameter-to-observable mapping), and $\bepsilon \in \RR^{n_y}$ the observation error. 
Under the Bayesian perspective, $\btheta$ is modeled as a random vector whose PDF represents the belief (i.e., uncertainty) about $\btheta$. When new data $\by$ is acquired from an experiment performed at design $\bd$, the PDF of $\btheta$ is updated via Bayes' rule:
\begin{align}
    p( \btheta | \by, \bd ) = \frac{p( \by | \btheta, \bd ) \, p(\btheta | \bd)}{p(\by | \bd)}
    \label{e:Bayes}
\end{align}
where $p(\btheta | \bd)$
is the prior PDF, $p( \by | \btheta, \bd )$ is the likelihood, $p(\by | \bd)$ is the marginal likelihood (model evidence), and $p(\btheta | \by, \bd)$ is the posterior PDF.
The prior thus depicts the uncertainty in $\btheta$ before seeing any data, and the posterior depicts the updated uncertainty after observing $\by$ from an experiment performed at $\bd$. 
We can reasonably assume that the prior is unchanged by the design alone, i.e., $p(\btheta|\bd) = p(\btheta)$. 
The likelihood corresponding to \cref{e:obs_model} can be evaluated through the PDF of $\bepsilon$: $p( \by | \btheta, \bd )=p_{\bepsilon}\left( \by - \mathbf{G}(\btheta, \bd) \right)$. 

\begin{figure}[htb]
    \centering
    \includegraphics[width=0.6\textwidth]{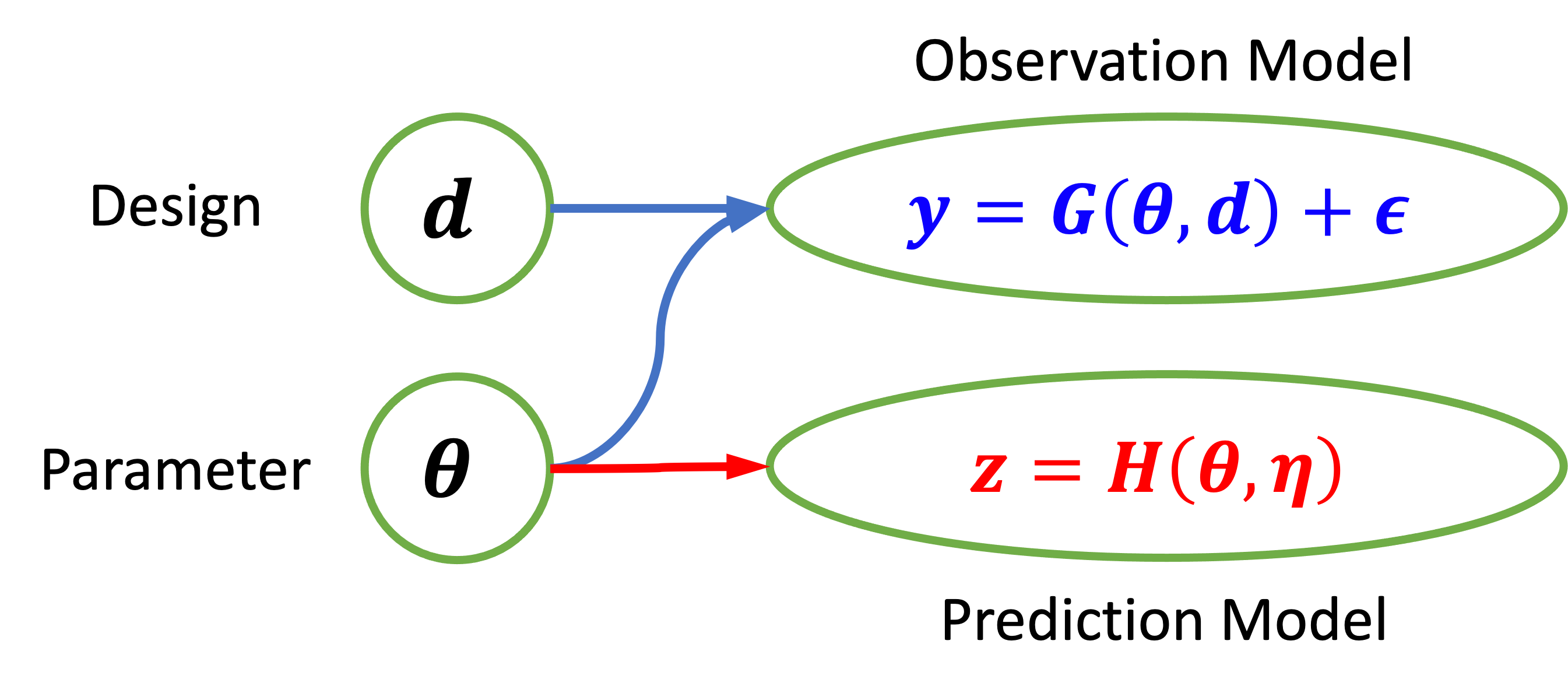}
    \caption{Overview of the relationships among different variables in the GO-OED framework.}
    \label{fig:GOOED_frame}
\end{figure}

Further consider a \emph{prediction model}
\begin{align}
    \bz = \mathbf{H}(\btheta, \boldsymbol{\eta})
    \label{e:pred_model}
\end{align}
where $\bz\in \CZ \subseteq \RR^{n_{z}}$ is the vector of predictive QoIs that depends on the model parameters $\btheta$ (but does not depend on the experimental design $\bd$) and prediction stochastic variable $\boldsymbol{\eta} \in \RR^{n_{\eta}}$, and 
$\mathbf{H}: \bTheta \times \RR^{n_{\eta}} \to \CZ$ 
is a nonlinear stochastic 
prediction forward model (parameter-to-QoI mapping). \Cref{fig:GOOED_frame} illustrates the relationship of $\btheta$, $\by$, $\bd$, and $\bz$ through the observation and prediction models.

The \emph{prior-predictive} and \emph{posterior-predictive} PDFs for $\bz$ are respectively
\begin{align}
p(\bz)&=\int_{\bTheta} p(\bz|\btheta) \, p(\btheta)\, \text{d}\btheta,\\
p(\bz|\by,\bd)&=\int_{\bTheta} p(\bz|\btheta) \, p(\btheta|\by,\bd)\, \text{d}\btheta.
\end{align}
If the prediction is deterministic $\bz = \mathbf{H}(\btheta)$ (i.e., no prediction stochasticity $\boldsymbol{\eta}$) then the respective PDFs are defined by pushforward probability measures under suitable conditions.\footnote{In situations where we can separate \cref{e:pred_model} into deterministic and stochastic parts, e.g.,  $\mathbf{H}(\btheta, \boldsymbol{\eta})=\mathbf{H}(\btheta)+\boldsymbol{\eta}$, then we make the distinction between \emph{prior-/posterior-predictive} and \emph{prior-/posterior-pushforward} PDFs.  ``Predictive'' refers to $\bz$, and ``pushforward'' refers to the deterministic prediction portion (e.g., $\mathbf{H}(\btheta)$). The two names coincide when the prediction model is deterministic: $\bz = \mathbf{H}(\btheta)$. }
Whether stochastic or deterministic, we proceed with using $\bz$ to denote the predictive QoIs.

We take a decision-theoretic approach~\cite{Lindley1956} to quantify the value of an experiment. 
Let $u(\bd,\by,\btheta)$ 
be the \emph{utility} when $\by$ is obtained from an experiment performed at design $\bd$ and the true data-generating parameters are $\btheta$. Since $\btheta$ and $\by$ are unknown when designing the experiment, we take their joint expectation to arrive at the \emph{expected utility}
\begin{align}
    U(\bd) = \int_{\CY}\int_{\bTheta} u(\bd, \by, \btheta) \, p(\btheta, \by | \bd) \,\text{d}\btheta \,\text{d}\by.
\label{e:EU}
\end{align}

In non-GO-OED, $u(\bd,\by,\btheta)$ is typically chosen to be the KL divergence from the $\btheta$-prior to the $\btheta$-posterior:
{\begin{align}
    u_{\btheta}(\bd, \by):= \DKL \left( p_{\btheta|\by,\bd} \,||\, p_{\btheta}  \right) = \int_{\mathcal{\bTheta}} p({\btheta}|\by, \bd) \ln \left[ \frac{p({\btheta} | \by, \bd)}{p({\btheta})} \right] \,\text{d}{\btheta}
    ,\label{e:non_GO_utility}
\end{align}
which happens to be independent of $\btheta$ since it is integrated out.
Substituting \cref{e:non_GO_utility} into \cref{e:EU} yields
\begin{align}
    U_{\btheta}(\bd) 
    &= \int_{\CY} \int_{\bTheta} p({\btheta}|\by, \bd) \ln \left[ \frac{p({\btheta} | \by, \bd)}{p({\btheta})} \right] p(\by | \bd) \, \text{d}{\btheta} \, \text{d}\by
    = \EE_{\by|\bd} \left[ \DKL(p_{\btheta|\by,\bd} \,||\, p_{\btheta}) \right]\label{e:non_GO_EU_KL} \\ 
    &= \int_{\CY} \int_{\bTheta} p({\btheta},\by| \bd) \ln \left[ \frac{p({\btheta} , \by| \bd)}{p(\by |\bd) \, p({\btheta})} \right] \,\text{d}{\btheta} \, \text{d}\by
    = \CI(\btheta;\by|\bd),\label{e:non_GO_EU_MI}
\end{align}
which is the expected KL divergence (or EIG) on the predictive QoIs $\btheta$ (per \cref{e:non_GO_EU_KL}) and also the MI between 
$\by$ and 
$\btheta$ conditioned on $\bd$ (per \cref{e:non_GO_EU_MI})~\cite{Lindley1956}.}

Since GO-OED now targets the predictive QoIs $\bz$, we follow~\cite{Bernardo1979} and analogously {choose $u(\bd,\by,\btheta)$ to be} the KL divergence from the
$\bz$-prior-predictive  to the $\bz$-posterior-predictive:
{
\begin{align}
    u_{\bz}(\bd, \by) := \DKL \left( p_{\bz|\by,\bd} \,||\, p_{\bz}  \right) 
    = \int_{\mathcal{Z}} p({\bz}|\by, \bd) \ln \left[ \frac{p({\bz} | \by, \bd)}{p({\bz})} \right] \,\text{d}{\bz}
    ,
\label{e:GO_utility}
\end{align}
which again happens to be independent of $\btheta$.}
Substituting \cref{e:GO_utility} into \cref{e:EU}, we obtain
\begin{align}
    U(\bd) 
    &= \int_{\CY} \int_{\mathcal{Z}} p({\bz}|\by, \bd) \ln \left[ \frac{p({\bz} | \by, \bd)}{p({\bz})} \right] p(\by | \bd) \, \text{d}{\bz} \, \text{d}\by
    = \EE_{\by|\bd} \left[ \DKL(p_{\bz|\by,\bd} \,||\, p_{\bz}) \right]\label{e:GO_EU_KL} \\ 
    &= \int_{\CY} \int_{\mathcal{Z}} p({\bz},\by| \bd) \ln \left[ \frac{p({\bz} , \by| \bd)}{p(\by |\bd) \, p({\bz})} \right] \,\text{d}{\bz} \, \text{d}\by
    = \CI(\bz;\by|\bd),\label{e:GO_EU_MI}
\end{align}
which is the expected KL divergence (or EIG) on the predictive QoIs $\bz$ (per \cref{e:GO_EU_KL}) and also the MI between 
$\by$ and 
$\bz$ conditioned on $\bd$ (per \cref{e:GO_EU_MI}).
Notably, as stated by Theorem 1 in~\cite{Bernardo1979},
\begin{align}
\CI(\bz;\by|\bd) \leq \CI(\btheta;\by|\bd)
 \label{e:z_bijective}
\end{align}
with equality if $\bz$ is a one-to-one transformation of $\btheta$. That is, the EIG (or MI) on the predictive QoIs $\bz$ is always less or equal than the EIG (or MI) on the parameters $\btheta$.

Lastly, solving the GO-OED problem entails finding the optimal design from a design space $\CD$ to maximize the expected utility:
\begin{align}
    \bd^{*} = \argmax_{\bd\in\CD} U(\bd).
    \label{e:GO_OED}
\end{align}
\section{Numerical Methods}
\label{s:methods}

\subsection{Monte Carlo Estimation of the Expected Utility}

In order to solve the GO-OED problem in \cref{e:GO_OED}, we need to be able to evaluate $U(\bd)$. However, $U(\bd)$ generally does not have a closed form and must be approximated numerically. We proceed to build a MC estimator for $U(\bd)$.

An initial MC estimator for \cref{e:GO_EU_KL} may be written as
\begin{align}
        U(\bd) 
        &= \int_{\CY} \int_{\mathcal{Z}} p({\bz}|\by, \bd) \ln \left[ \frac{p({\bz} | \by, \bd)}{p({\bz})} \right] p(\by | \bd) \, \text{d}{\bz} \, \text{d}\by  
        \nonumber\\
        &= \int_{\CY} \int_{\mathcal{Z}}  \ln \left[ \frac{p({\bz} | \by, \bd)}{p({\bz})} \right] p(\by, {\bz} | \bd)\, \text{d}{\bz} \, \text{d}\by  
        \nonumber\\
        &= \int_{\CY} \int_{\mathcal{Z}}  \int_{\bTheta} \ln \left[ \frac{p({\bz} | \by, \bd)}{p({\bz})} \right]  p(\btheta, \by, {\bz} | \bd) \, \text{d}\btheta \, \text{d}{\bz} \, \text{d}\by  
        \label{e:MC0}\\
        &= \int_{\CY} \int_{\mathcal{Z}}  \int_{\bTheta} \ln \left[ \frac{p({\bz} | \by, \bd)}{p({\bz})} \right]  p(\btheta) \, p(\by | \btheta, \bd) \, p({\bz} | \btheta) \, \text{d}\btheta \, \text{d}{\bz} \, \text{d}\by
        \nonumber\\
        &\approx \frac{1}{n} \displaystyle\sum_{i=1}^{n} \left\{ \ln{\left[ p(\bz^{(i)} | \by^{(i)}, \bd) \right]}  - \ln{\left[ p(\bz^{(i)}) \right]}  \right\}, \label{e:MC1}
\end{align}
where 
we can sample 
$\btheta^{(i)}\sim p(\btheta)$, $\by^{(i)}\sim  p(\by|\btheta^{(i)},\bd)$ (using observation model \cref{e:obs_model}), $\bz^{(i)}\sim p(\bz|\btheta^{(i)})$ (using prediction model \cref{e:pred_model}).
We do not explicitly have the prior-predictive PDF $p(\bz^{(i)})$, but using MC samples we can estimate this density.
However, we cannot evaluate the posterior-predictive PDF $p(\bz^{(i)} | \by^{(i)}, \bd)$ in the first term of \cref{e:MC1}, nor can we directly estimate this conditional density from MC samples since we do not have multiple $\bz$ samples for each $\by^{(i)}$. 
This motivates the following {NMC} estimator, continuing from~\cref{e:MC0}:
\begin{align}
    U(\bd) &= \int_{\CY} \int_{\mathcal{Z}}  \int_{\bTheta} \ln \left[ \frac{p({\bz} | \by, \bd)}{p({\bz})} \right]  p(\btheta, \by, {\bz} | \bd) \, \text{d}\btheta \, \text{d}{\bz} \, \text{d}\by 
    \nonumber\\
    &= \int_{\CY}  \int_{\bTheta} \int_{\mathcal{Z}} \ln \left[ \frac{p({\bz} | \by, \bd)}{p({\bz})} \right]  p(\by|\bd) \, p(\btheta | \by, \bd) \, p({\bz} | \btheta) \, \text{d}\btheta \, \text{d}{\bz} \, \text{d}\by
    \nonumber\\
    &\approx \frac{1}{ n_{\rm{out}}} \displaystyle\sum_{i=1}^{n_{\rm{out}}} \left\{ \frac{1}{n_{\rm{in}} } \sum_{j=1}^{n_{\rm{in}}} \ln\left[ p(\bz^{(i,j)} | \by^{(i)}, \bd) \right]  - \ln\left[ p(\bz^{(i)}) \right]   \right\}
    \label{equ:MC_expected}
\end{align}
where $n_{\rm{out}}$ and $n_{\rm{in}}$ are respectively the number of samples for the outer and inner loops. 
For the first term, the outer loop is tasked with sampling $\by^{(i)}\sim  p(\by|\bd)$, which is achieved by sampling $\tilde{\btheta}^{(i)}\sim p(\btheta)$ followed by $\by^{(i)}\sim  p(\by|\tilde{\btheta}^{(i)},\bd)$. 
The inner loop then samples $\btheta^{(i,j)}\sim p(\btheta|\by^{(i)},\bd)$ and $\bz^{(i,j)}\sim p(\bz|
\btheta^{(i,j)})$.
For the second term, since $\ln[p(\bz)]$ is independent from both $\by$ and $\bd$, the term may be estimated separately from the above loops. Here we write its estimation under the outer loop so to use the same $\tilde{\btheta}^{(i)}$ samples to generate the prior-predictive $\bz^{(i)}$ samples, but noting that this is done only once at the beginning of the $\bd$ optimization and need not be repeated at each $\bd$ (see \cref{alg:GO_OED}). 
{Furthermore, if one is interested only in identifying $\bd^{\ast}$, then the $\ln\left[ p(\bz^{(i)}) \right]$ term can be omitted altogether, as it is independent of $\bd$ and does not influence the optimal design location. However, if one is also interested in the EIG values for various designs, the term should be retained.}

The key of the {NMC} structure is that it allows multiple $\bz^{(i,j)}$ samples to be generated for each $\by^{(i)}$, and hence density estimation can be carried out for $p(\bz | \by^{(i)}, \bd)$. Furthermore, while the prior, likelihood, and predictive sampling all can be done easily, $\btheta^{(i,j)}\sim p(\btheta|\by^{(i)},\bd)$ is posterior sampling and non-trivial. In the next section, we discuss our approach to these two numerical challenges in the MC estimator for $U(\bd)$: posterior sampling and density estimation.

\subsection{Posterior Sampling and Density Estimation}
\label{ss:MCMC_KDE}

\subsubsection{Markov Chain Monte Carlo for Sampling $p(\btheta|\by^{(i)},\bd)$}
\label{sss:MCMC}

We elect to perform posterior sampling via MCMC, although the overall GO-OED framework is agnostic to the choice of posterior sampling method. MCMC will asymptotically sample the posterior distribution under sufficient conditions of ergodicity and detailed balance. 
In particular, we adopt an existing parallel Python MCMC implementation called \emph{emcee}~\cite{foreman-mackey_emcee_2013}.
Emcee is built upon the 
affine invariant 
stretch move algorithm \cite{goodman_ensemble_2010} that is designed to perform 
well under 
all 
linear transformations and therefore insensitive to correlation among parameters. 
The algorithm involves an ensemble of $n_w$ walkers $S=\{\btheta_1, \btheta_2, \cdots, \btheta_{n_w}\}$. For the $i$th walker currently at 
$\btheta_i$, 
the proposal procedure first randomly samples a $\btheta_j$ among the current positions of the other $(n_w-1)$ walkers and then computes
\begin{align}
    \btheta_p = \btheta_j + \gamma \left( \btheta_i - \btheta_j \right)
    \label{equ:EMCEE}
\end{align}
where $\gamma$ is drawn from a PDF $p(\gamma)$ satisfying
$    p\left(\frac{1}{\gamma}\right) = \gamma p(\gamma)$
in order to achieve the affine invariant property. For example~\cite{goodman_ensemble_2010} uses
\begin{align}
    p(\gamma) \propto \left\{ 
    \begin{array}{cc}
    \frac{1}{\sqrt{\gamma}} & \text{if } \gamma\in\left[ \frac{1}{a}, a \right] \\
    0 & \text{otherwise}
    \end{array}
    \right.
\end{align}
where $a>1$ is an adjustable hyperparameter.
Detailed balance is satisfied if the proposed point is accepted with probability
\begin{align}
    \alpha = \min\left(1, \gamma^{n_{\theta}-1} \frac{p(\btheta_p|\by,\bd)}{p\(\btheta_i|\by,\bd\)} \right)
\end{align}
where the posterior density ratio can be calculated from just the prior and likelihood (marginal likelihood terms cancel out in the ratio).
The emcee package further parallelizes the algorithm by splitting all walkers into two subsets,
where walkers from within each subset are then updated by proposing from the other subset. 
The use of multiple walkers can potentially provide better exploration of multi-modal posteriors.

We note that in the GO-OED context, the posterior sampling is done conditioning on each $\by^{(i)}$ sample, and we always have access to the true $\tilde{\boldsymbol{\theta}}^{(i)}$ that generated this $\by^{(i)}$ (see the sampling procedure described just below \cref{equ:MC_expected}). Hence, MCMC can be accelerated by initializing the walkers at $\tilde{\boldsymbol{\theta}}^{(i)}$, which corresponds to an ``oracle estimator'' that is generally much closer to the center region of the posterior than a randomly initialized point from the prior. Thus, MCMC in GO-OED is more benign, and requires fewer iterations, than a typical Bayesian inference problem. However, for unidentifiable and multi-modal posteriors, this technique is less effective and samplers suited for multi-modality may be more appropriate~\cite{earl2005parallel, latz2021generalized, pompe2018framework, paulin2019error, catanach2018bayesian}. As with general MCMC, convergence diagnostics should be evaluated to make sure the choice of sampler and its parameters are amenable to the problem~\cite{cowles1996markov, roy2020convergence}.

\subsubsection{Kernel Density Estimation for  $p(\bz)$ and $p(\bz|\by^{(i)},\bd)$}
\label{sss:KDE}

We perform density estimation for the prior- and posterior-predictive PDFs, respectively $p(\bz)$ and $p(\bz | \by^{(i)}, \bd)$, using KDE. 
Given samples $(\bz_1,\bz_2,\cdots,\bz_n)$, the estimated PDF from KDE is
\begin{align}
    \hat{p}_{{\bB}}(\bz) = \frac{1}{n}\displaystyle\sum_{i=1}^{n}K_{{\bB}} (\bz-\bz_i) 
\end{align}
where $K_{{\bB}}$ is the kernel 
and ${\bB}$ is its bandwidth matrix.  
We use the Scikit-Learn KDE implementation
\cite{scikit-learn} that
employs an isotropic Gaussian kernel with ${\bB=\mathrm{diag}(b^2,b^2,\ldots,b^2)}$: 
\begin{align}
    K_{{\bB}} (\bz-\bz_i) = \prod_{j=1}^{n_z}\frac{1}{\sqrt{2\pi} {b}}\exp\left[-\frac{(\bz_j-\bz_{ij})^2}{2{b}^2}\right].
\end{align}
Thus, {$b$} plays the role of standard deviation of the kernel, and smaller {$b$} leads to sharper peaks around each sample while larger {$b$} induces a more diffusive effect. 
{When applying KDE to $p(\bz | \by^{(i)}, \bd)$, the optimal choice of $b$ may vary for each $\by^{(i)}$ sample and every new $\bd$. While various methods for selecting {$b$} have been proposed \cite{doi:10.1080/01621459.1990.10475307, cao_comparative_1994, jones_brief_1996}, they become computationally taxing if these calculations must be repeated for 
every $\by^{(i)}$ sample and each new $\bd$. 
Instead,} to maintain a reasonable computational cost for bandwidth selection, at each $\bd$, we perform 5-fold cross-validation to select the optimal {$b$} during the first few outer-loop iterations, 
and then choose the average value of those optimal {$b$}'s and fix it in all subsequent computations for that $\bd$.
Bandwidth can certainly affect the estimation of $U(\bd)$, where too small of a bandwidth produces sharper posteriors and therefore higher perceived EIG and overestimate of $U(\bd)$,  while too large a bandwidth tends to underestimate $U(\bd)$. We will explore the effects of KDE bandwidth in the numerical experiments.

\subsection{Bayesian Optimization}
\label{ss:BO}

Lastly, we need an optimization algorithm to solve \cref{e:GO_OED}. Such algorithm needs to handle noisy objectives since only MC estimates of $U(\bd)$ are available. Grid search may be feasible for low dimensional $\bd$ (e.g., $n_d \leq 3$) but the number of grid points grows exponentially with $n_d$.
While previous efforts have investigated derivative-free (e.g., nonlinear simplex and simultaneous perturbation stochastic approximation)~\cite{Huan2013} and gradient-based (e.g., Robbins-Monro and sample average approximation)~\cite{Huan2014} optimization methods in the context of non-GO-OED, the former can be slow in convergence
and the latter requires gradient access and more prone to local optima.

We explore the use of BO in this work favoring its globally convergent properties, 
high sample efficiency and noise tolerance, and suitability for expensive objective function evaluations.
In particular, we adopt the Nogueira Python BO package~\cite{fernando2014BO}. 
BO describes the objective function $U$ as a Gaussian process (GP)~\cite{Rasmussen2006,Gramacy2020}, and its GP prior can 
be updated to a posterior analytically when new ``observations'' (i.e., evaluations) of $U$ become available. 
The next evaluation point for $U$ is then determined by optimizing a relevant acquisition function derived from the GP, which does not require gradient of $U$. 
We summarize below the three main BO steps along with our choice of algorithm settings.

\subsubsection{Step 1: Gaussian Process Regression}
\label{sss:BO1}

BO uses a GP to represent the uncertainty of a random function $U: \CD \to \RR$ that we wish to optimize. 
The GP is specified by a mean function $m(\bd)$ and covariance kernel $k(\bd,\bd')$.
While different kernel choices are possible, we adopt the Mat\'{e}rn kernel~\cite{Rasmussen2006}, a popular choice that also generalizes the radial basis function:
\begin{align}
    k(\bd,\bd') = \frac{1}{\Gamma(\nu)2^{\nu-1}} \left(\frac{\sqrt{2\nu}}{l} {\norm{\bd-\bd'}{2}}\right)^\nu B_\nu \left(\frac{\sqrt{2\nu}}{l} {\norm{\bd-\bd'}{2}}\right)
\end{align}
where $\nu$ controls the smoothness, $l$ is the length scale, 
$\Gamma$ is the gamma function, and $B_\nu$ is the modified Bessel function. {For simplicity, we fix $l=1$ and $\nu=2.5$ to their default values in the package. However, these hyperparameters certainly can be re-tuned, such as after every few updates of the GP, to further enhance performance.}
As new evaluations of the random function $\bU=[U^{(1)},\dots,U^{(k)}]^{\top}$ at locations $\bD=[\bd^{(1)},\dots,\bd^{(k)}]^{\top}$ become available, the GP's prior mean and covariance are updated to the posterior mean $m(\bd|\{\bD,\bU\})$ and covariance $\Cov[U(\bd),U(\bd')|\{\bD,\bU\}]$ following standard GP regression update formulas~\cite{Rasmussen2006}.

\subsubsection{Step 2: Acquisition Function}
\label{sss:BO2}

{The selection of the next location, $\bd^{(k+1)}$, to evaluate $U$ is guided by a real-valued \emph{acquisition function}, denoted as $a(\bd)$. At each BO iteration, the acquisition function is updated using the latest posterior GP
and subsequently optimized to determine the next evaluation point:
\begin{align}
\bd^{(k+1)}=\argmax_{\bd\in\CD} a(\bd).
\end{align}
The acquisition function serves as a heuristic to guide the BO algorithm towards regions with high potential for containing the optimum of the optimization problem in \cref{e:GO_OED}.
Since the acquisition function must be optimized at every iteration, it needs to be computationally inexpensive to evaluate.}

{Various acquisition function have been proposed in the BO literature, including the upper confidence bound (UCB), probability of improvement (PI), and expected improvement (EI) (see, e.g.,~\cite{jones1998efficient,movckus1975bayesian,zhan2020expected}). 
In our work, we adopt a 99.5\% UCB, defined as
\begin{align}
    a(\bd) = m(\bd|\{\bD,\bU\}) + \kappa \,\sigma(\bd|\{\bD,\bU\}),
\end{align}
where $\kappa= 2.56$ is fixed to correspond to a 99.5\% confidence level of the GP, and $\sigma(\bd|\{\bD,\bU\}):=\sqrt{\Cov[U(\bd),U(\bd)|\{\bD,\bU\}]}$ is the GP posterior standard deviation.
A high value of $a(\bd)$ thus reflects a combination of high mean value (i.e., to exploit the current GP estimate) and high standard deviation (i.e., to explore regions with high GP uncertainty). 
}

\subsubsection{Step 3: Optimization Update}
\label{sss:BO3}

The next location $\bd^{(k+1)}$ to evaluate $U$ is then selected by maximizing the acquisition function.
This inner optimization subproblem is inexpensive by design and needs not be solved very accurately in practice. We solve it using the limited-memory Broyden-Fletcher-Goldfarb-Shanno algorithm with box constraints (L-BFGS-B)~\cite{Nocedal2006}[Ch. 7] and with multiple restarts. 
Upon obtaining $U(\bd^{(k+1)})$, the GP can then be updated via Step 1. The cycle can be repeated until we reach a stopping criterion (e.g., maximum allowable iteration, lack of improvement to the highest value of $U(\bd)$ encountered).

\subsection{Summary of the Overall Algorithm}

Pseudocode for our overall MCMC-based GO-OED algorithm is provided in \cref{alg:GO_OED}.

\begin{algorithm}[htb]
	\caption{MCMC-based GO-OED.}
    \label{alg:GO_OED}
	\begin{algorithmic}[1]
		\REQUIRE Prior $p(\btheta)$; observation forward model $G(\btheta, \bd)$; likelihood $p(\by|\btheta,\bd)$; prediction model $H(\btheta,\boldsymbol{\eta})$; predictive-likelihood $p(\bz|\btheta)$; MC sample size $n_{\text{out}}$, $n_{\text{in}}$; initial design $\bd_0$; emcee hyperparameters $n_w$, $a$; KDE kernel hyperparameter $h$; BO hyperparameters $\nu$, $l$, termination criteria;
        \STATE 
        Draw $n_{\text{out}}$ prior samples ${\btheta}^{(i)}\sim p({\btheta})$ and prior-predictive samples $\bz^{(i)}\sim p(\bz|\btheta^{(i)})$, estimate prior-predictive PDF $p(\bz)$ using KDE (\cref{sss:KDE}){\textsuperscript{\textdagger}};
        \STATE Set $k=0$, initial $\bd_{0}$;
		\WHILE{BO termination criteria not met}
        \FOR{$i=1,\ldots,n_{\text{out}}$}
        \STATE Sample $\tilde{\btheta}^{(i)}\sim p({\btheta})$,
        $\by^{(i)}\sim  p(\by|\tilde{\btheta}^{(i)},\bd_{k})$;
        \FOR{$j=1,\ldots,n_{\text{in}}$}
        \STATE Sample posterior $\btheta^{(i,j)}\sim p(\btheta|\by^{(i)},\bd)$ via MCMC (\cref{sss:MCMC});
        \STATE Sample posterior-predictive $\bz^{(i,j)}\sim p(\bz|\btheta^{(i,j)})$; 
        \STATE Estimate posterior-predictive PDF $p(\bz|\by^{(i)},\bd)$ using KDE (\cref{sss:KDE}); 
        \ENDFOR
        \ENDFOR
        \STATE Estimate $U(\bd_k)\approx \frac{1}{ n_{\rm{out}}} \displaystyle\sum_{i=1}^{n_{\rm{out}}} \left\{ \frac{1}{n_{\rm{in}}} \sum_{j=1}^{n_{\rm{in}}}  \ln\left[ p(\bz^{(i,j)} | \by^{(i)}, {\bd_k}) \right]   - \ln\left[ p(\bz^{(i)}) \right]  \right\} $ (see \cref{equ:MC_expected}){\textsuperscript{\textdagger}};
        \STATE Update GP mean and covariance (\cref{sss:BO1});
        \STATE Optimize updated GP acquisition function to identify next point $\bd_{k+1}$ to evaluate objective (\cref{sss:BO2,sss:BO3});
        \STATE $k=k+1$;
		\ENDWHILE
    \STATE Return $\bd_k$ as the numerical estimate for $\bd^{\ast}$, and  ${U}(\bd_k)$ for ${U}(\bd^{\ast})$;
	\end{algorithmic}
    {\textsuperscript{\textdagger} Note: if only seeking $\bd^{\ast}$ and not explicitly the EIG values, Step 1 may be skipped, and Step 12 can omit the $\ln\left[ p(\bz^{(i)}) \right]$ term because it does not depend on $\bd$.}
\end{algorithm}

\section{Numerical Experiments and Results}
\label{s:results}

We present a series of numerical experiments with increasing complexity and designed to illuminate different aspects of the GO-OED approach. \Cref{ss:1D} starts with several cases with one-dimensional (1D) parameter, observation, and design spaces. The primary purposes of these examples are to validate GO-OED by comparing it with accessible and accurate non-GO-OED results under situations where theory shows they should agree, and to explore GO-OED's numerical behavior (e.g., effect of KDE bandwidth). \Cref{ss:2D} increases to two-dimensional (2D) parameter, observation, and design spaces, {demonstrating} the effectiveness of BO. {\Cref{ss:ND} further extends to multi-dimensional spaces, focusing on computational cost as sample sizes and dimensionality vary.} Lastly, \cref{ss:CD} presents a problem of sensor placement in a convection-diffusion field with predictive QoIs that include future concentrations at various locations and flux across a boundary. The last problem involves even higher dimensional settings and provides illustrations that involves physics-based modeling. 
{Unless stated otherwise, all numerical examples use sample sizes of  $n_{\text{in}} =1000$ and $n_{\text{out}} =1000$.}

\subsection{1D Nonlinear Test Problems}
\label{ss:1D}

We begin with a 1D test problem from~\cite{Huan2013} to explore some basic properties of our GO-OED approach. Consider a nonlinear observation model:
\begin{align}
    y(\theta, d) &= G(\theta, d) + \epsilon \nonumber\\
    &= \theta^3d^2 + \theta \exp\left( -|0.2-d| \right) + \epsilon,
    \label{equ:obs1}
\end{align}
where all variables are scalar, $d\in [0,1]$, $\epsilon\sim\mathcal{N}(0,10^{-4})$, and prior $\theta\sim \CU(0,1)$. With this observation model, below we present cases involving different combinations of prediction models.

\paragraph{Case {BM}} First, we set up a benchmark ({BM}) case where the prediction model is $\theta$ itself:
\begin{align}
    z = H(\theta) = \theta.
\end{align}
GO-OED therefore coincides with non-GO-OED and near-identical results are expected. 
As a high-accuracy reference solution, we estimate $U(d)$ using a gridding method on a fine grid of 2000 nodes for discretizing the parameter space $\bTheta=[0,1]$. The unnormalized posterior PDF (i.e., likelihood times prior) is calculated on this grid and then normalized by approximating the marginal likelihood through the mid-point integration rule. 
However, gridding is not scalable to higher dimensional $\theta$ and only implementable for prediction models of $z=\theta$, therefore it is {not available} for other test cases.

\begin{figure}[htbp]
  \centering
  \subfloat[Expected utility]{
  \label{fig:BM_all}
  \includegraphics[width=0.46\textwidth]{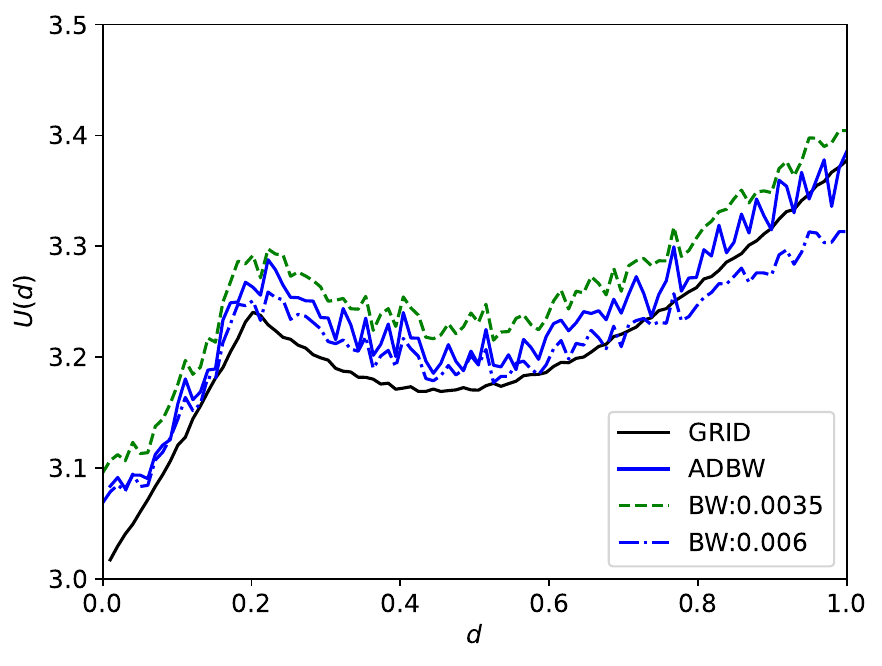}
  }%
  \subfloat[Optimized bandwidth values from ADBW]{
  \label{fig:BM_BW}
  \includegraphics[width=0.48\textwidth]{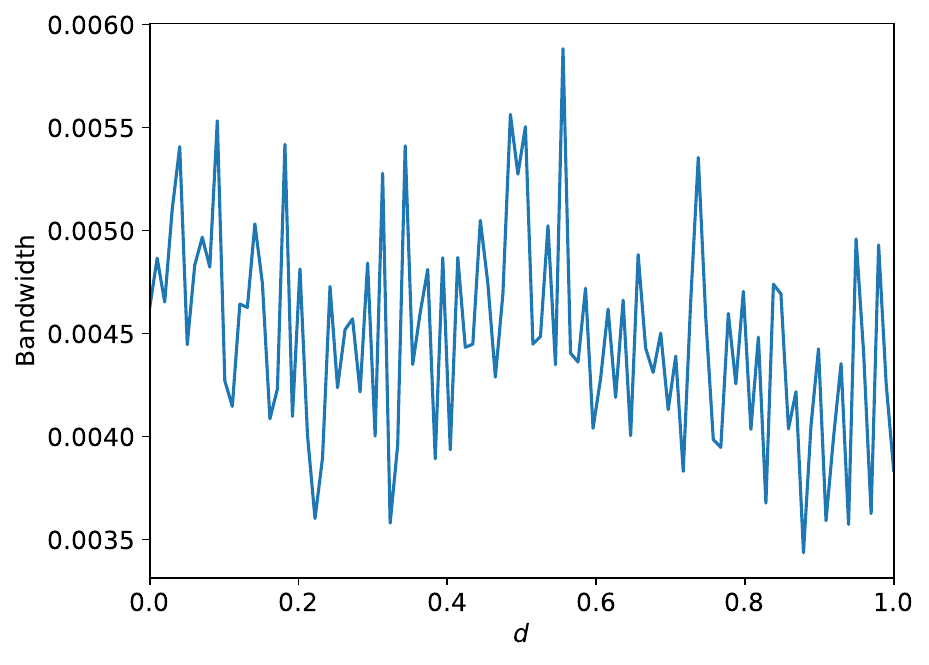}
  }  
  \caption{Case {BM}: expected utility (left) and optimized KDE bandwidth in ADBW across $d$ (right). GRID uses the gridding method for discretizing $\bTheta$ and is treated as the reference solution. ADBW and BW are the GO-OED estimators proposed in this paper, where ADBW uses adaptive bandwidth and BW uses fixed bandwidth. All methods agree on the general trends although bias and variance are noticeable for the three GO-OED methods.}
\end{figure}

\Cref{fig:BM_all} shows the expected utility computed using gridding (GRID), adaptive bandwidth (ADBW) (KDE bandwidth adaptively optimized at each $d$ using cross-validation as described in \cref{sss:KDE}), and fixed bandwidth (BW) (KDE bandwidth fixed across all $d$, illustrated at two pre-selected values) methods. MCMC used for ADBW and BW employed 1000 iterations with an additional 50 for burn-in. While all methods preserve the general trend in the expected utility, the discrepancies of ADBW and BW compared to the GRID reference is noticeable. Both ADBW and BW exhibit variance and bias, which result from the underlying MCMC and KDE computations. 
The optimized bandwidths for ADBW are shown in \cref{fig:BM_BW} across $d$, varying roughly between $0.0035$ and $0.006$. As shown in \cref{fig:BM_all}, the expected utility when setting BW to roughly ADBW's lower (0.0035) and upper (0.006) bandwidth limits approximately creates an envelopes around the ADBW expected utility curve. This is consistent with the anticipated behavior described in \cref{sss:KDE}, where a lower bandwidth tends to form narrower posterior distributions and overestimate the EIG, and vice versa for higher bandwidth. Overall, ADBW and BW perform similarly, suggesting that fixing bandwidth across the design space may be sufficient in practice. 

\begin{figure}[htbp]
  \centering
  \subfloat[Expected utility]{
  \label{fig:T1_all}
  \includegraphics[width=0.46\textwidth]{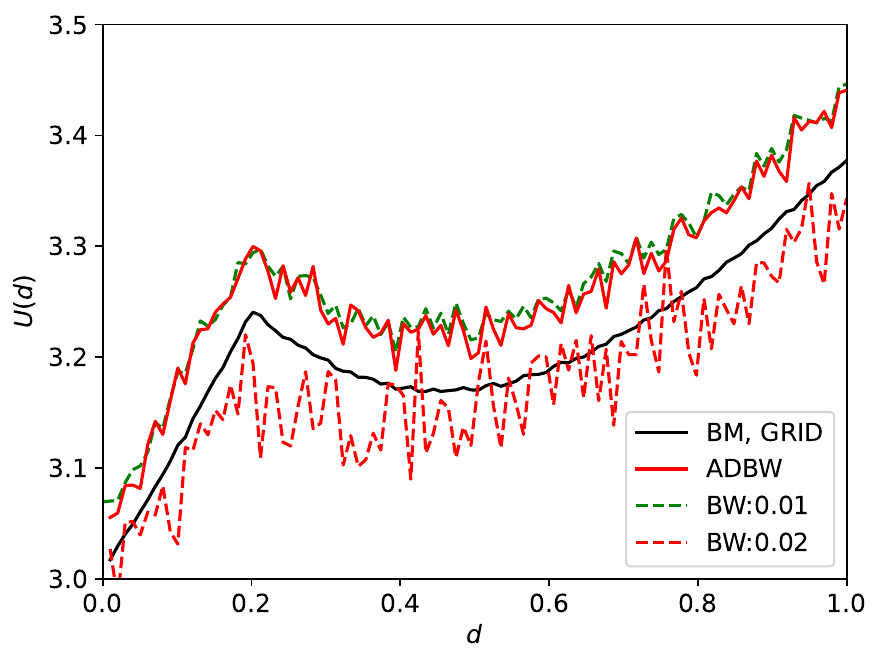}
  }%
  \subfloat[Optimized bandwidth values from ADBW]{
  \label{fig:T1_BW}
  \includegraphics[width=0.48\textwidth]{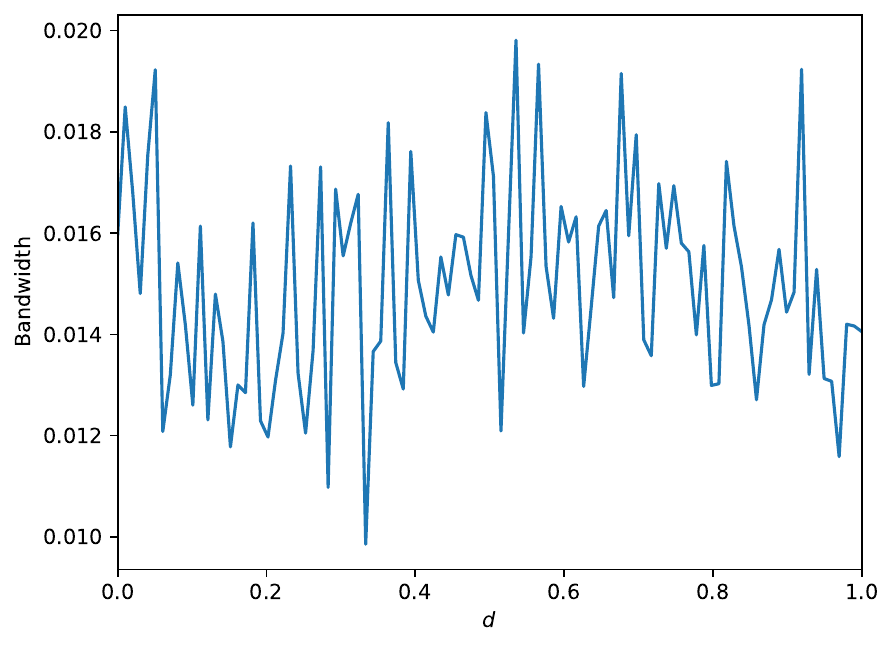}
  }
  \caption{Case T1: expected utility (left) and optimized KDE bandwidth in ADBW across $d$ (right). 
  A low bandwidth leads to an overestimated EIG, a high bandwidth leads to an underestimated EIG.
  }
\end{figure}

\paragraph{Case T1} 
In case T1, we adopt a nonlinear prediction model that differs from the observation model:
\begin{align}
    z = H(\theta) = \sin\theta + \theta \exp{(\theta + \abs{0.5-\theta})}.
    \label{equ:PredT1}
\end{align}
Since the prior for $\theta$ has compact support between $[0,1]$, the predictive QoI is bijective for the parameter within this range. From \cref{e:z_bijective}, a bijective parameter-to-observation mapping should lead to the same EIG as the non-GO-OED case just as in Case {BM}. 
This is confirmed by the expected utility curves shown in \cref{fig:T1_all}, all agreeing with the trend of the non-GO-OED GRID results in Case {BM} (which is equivalent to this case). The GO-OED methods here, however, exhibit additional variance and bias from the KDE and MCMC compared to those in the {BM} case. This is likely due to the increased sensitivity and variability of the mapping from $\theta$ to $z$.
The optimized bandwidths from ADBW is shown in \cref{fig:T1_BW} across $d$, 
varying roughly between $0.01$ and $0.02$. Similar to the previous case, in \cref{fig:T1_all} we observe that the expected utility curves when setting BW to around ADBW's lower (0.01) and upper (0.02) bandwidth limits create an envelope around ADBW, although the upper curve (from the lower bandwidth limit) is quite tight. Given the challenges in identifying a suitable fixed bandwidth \textit{a priori}, we will focus on the ADBW for upcoming cases. 

\paragraph{Case T2 and T3} 

\begin{figure}[htbp]
    \centering
    \includegraphics[width=0.6\textwidth]{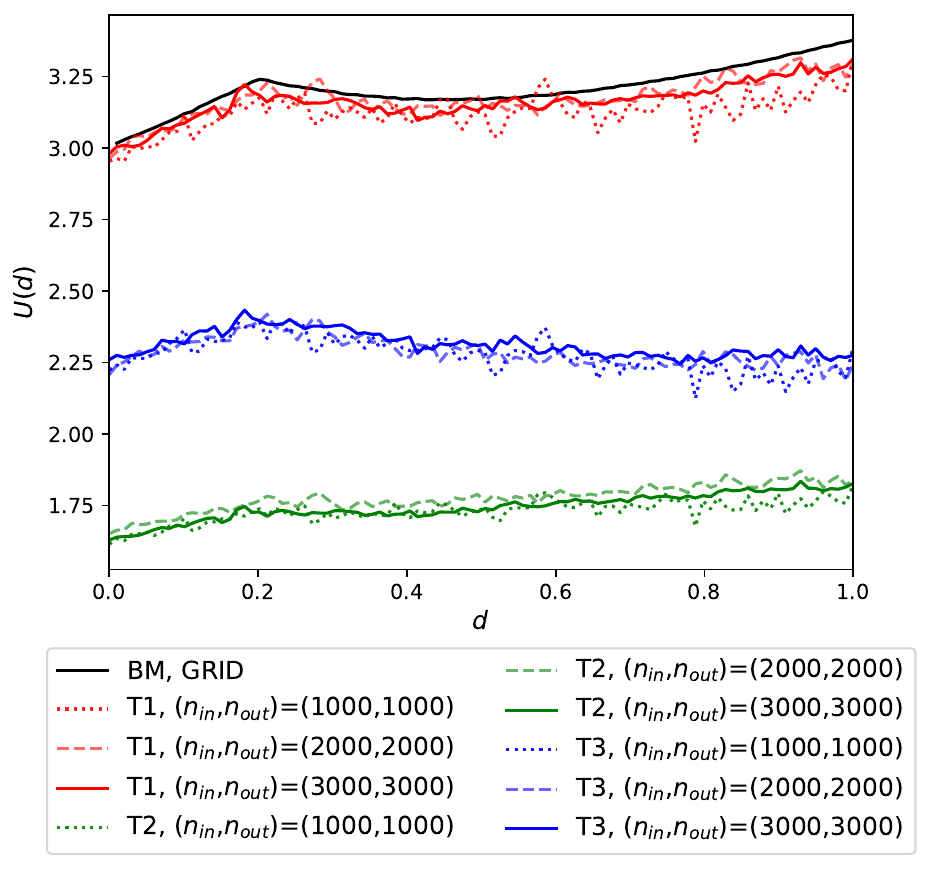}
    \caption{Case {BM}, T1, T2, T3: expected utility comparisons. The benchmark case ({BM}, GRID) has $z=\theta$ and therefore equal to the parameter EIG. Case T1 has a nonlinear but bijective mapping from the parameter to the QoI and so has the same EIG as {BM}.  Cases T2 and T3 are non-bijective QoIs and their EIGs are lower compared to the parameter EIG, per~\cref{e:z_bijective}. 
    {Cases using the proposed NMC estimator each has three plot lines, corresponding to EIG estimates under different sample sizes: $(n_{\text{out}},n_{\text{in}}) = (1000,1000), (2000,2000), (3000,3000)$.}}
    \label{fig:T2T3}
\end{figure}

In cases T2 and T3, we adopt observation models where $z$ is no longer bijective with $\theta$. 
These illustrations will show that reducing the uncertainty of parameters $\theta$ is not equivalent to reducing the uncertainty of predictive QoI $z$, leading to different optimal designs. 
Consider prediction model T2:
\begin{align}
    z={H}(\theta) &= \left\{
    \begin{aligned}
        &-100\theta+25, \ 0 \leq \theta < 0.15 \\
        &5, \ 0.15 \leq \theta \leq 0.7 \\
        &50\theta + 25, \ 0.7 < \theta \leq 1.0
    \end{aligned}
    \right. \label{equ:PredT2}
\end{align}
and prediction model T3:
\begin{align}
    z={H}(\theta) &= \frac{1}{\sqrt{2\pi}\sigma} \exp{\left[-\frac{(\theta-\mu)^2}{2\sigma^2}\right]} \quad \text{with} \quad \mu=0.3, \ \sigma=0.2. \label{equ:PredT3}
\end{align}
The expected utility curves for T2 and T3 are shown in \cref{fig:T2T3}, which are significantly different from that of {BM} and T1. Notably, T2 continuously increases with $d$ roughly at a constant rate and reaches an optimal around $d=1.0$, and T3 does not ``tail up'' as much as T1/{BM} when $d$ approaches 1.0 while reaching an optimal around $d=0.2$.
{The figure also includes plot lines under different sample sizes: $(n_{\text{out}},n_{\text{in}}) = (1000,1000), (2000,2000), (3000,3000)$. The lines with higher sample sizes exhibit lower variation, indicating a stabilization of the EIG estimation with increasing sample size.
All lines for each case preserve the same general trend, and their optimal design locations appear consistent with each other.}

\begin{figure}[htbp]
    \centering
    \subfloat[Posterior distributions conditioned on $y$ simulated at $\theta=0.1$
    ]{
    \includegraphics[width=0.46\textwidth]{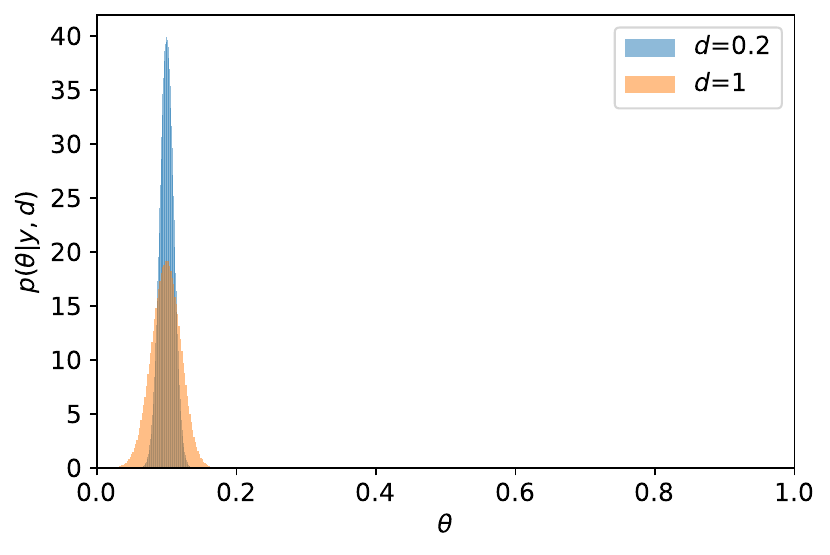}
    \label{fig:sample_posteriors_Param1}
    }\hfill
    \subfloat[Posterior distributions conditioned on $y$ simulated at $\theta=0.9$
    ]{
    \includegraphics[width=0.46\textwidth]{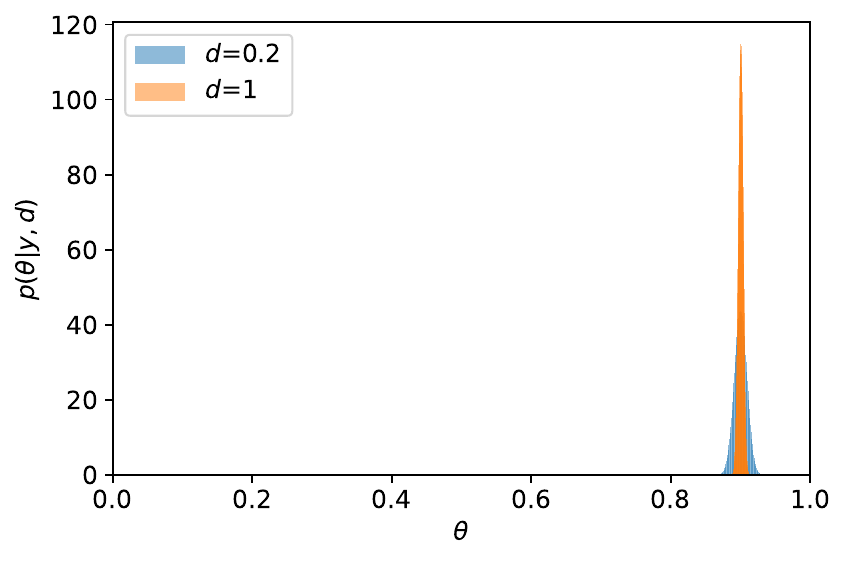}
    \label{fig:sample_posteriors_Param2}
    }%
    \caption{Case T3: example posterior distributions. The left plot conditions on $y$ simulated at $\theta=0.1$, and $d=0.2$ yields a narrower posterior; the right plot conditions on $y$ simulated at  $\theta=0.9$, and $d=1.0$ yields a narrower posterior.
    The variability in information gain thus can alter the ranking of the two designs. 
    When repeated for many samples of $\theta$ and $y$ and taking the expectation, it is the \emph{expected} information gain on $\theta$ that is higher at $d=1.0$ than at $d=0.2$. }
    \label{fig:sample_posteriors}
\end{figure}

To further illuminate the optimal design behavior of non-GO-OED versus GO-OED, we produce in \cref{fig:sample_posteriors} example posterior distributions when $d=0.2$ 
(T3 GO-OED optimum) 
and $d=1.0$ 
(non-GO-OED optimum). These examples are demonstrated by conditioning on $y$ simulated at $\theta=0.1$ and $\theta=0.9$ for parameter variety (recall prior $\theta\sim\CU(0,1)$). From \cref{fig:sample_posteriors_Param1} where the $y$ is simulated at $\theta=0.1$, $d=0.2$ yields a narrower posterior (KL divergence from prior to posterior, or information gain, of 3.15) while $d=1.0$ offers a slightly wider posterior (information gain 2.44).
\Cref{fig:sample_posteriors_Param2} presents another example for when $y$ is simulated at $\theta=0.9$, where the opposite is observed (information gain 3.89 at $d=1.0$ is higher than information gain 3.23 at $d=0.2$).
The variability in information gain thus can alter the ranking of the two designs. 
When this procedure is repeated for many  samples of $\theta$ and $y$ and taking the expectation, it is the \emph{expected} information gain on $\theta$ that becomes higher at $d=1.0$ than at $d=0.2$. 

\begin{figure}[htbp]
    \centering
    \subfloat[Prior-predictive distribution for T3 prediction model]{
    \includegraphics[width=0.47\textwidth]{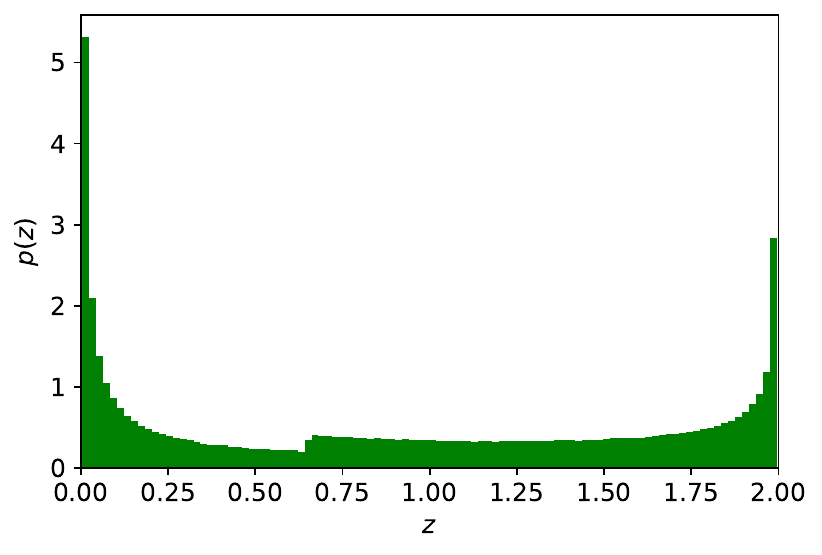}
    \label{fig:prior_predictive}
    }
    \\
    \subfloat[Posterior-predictive distributions conditioned on $y$ simulated at $\theta=0.1$
    ]{
    \includegraphics[width=0.46\textwidth]{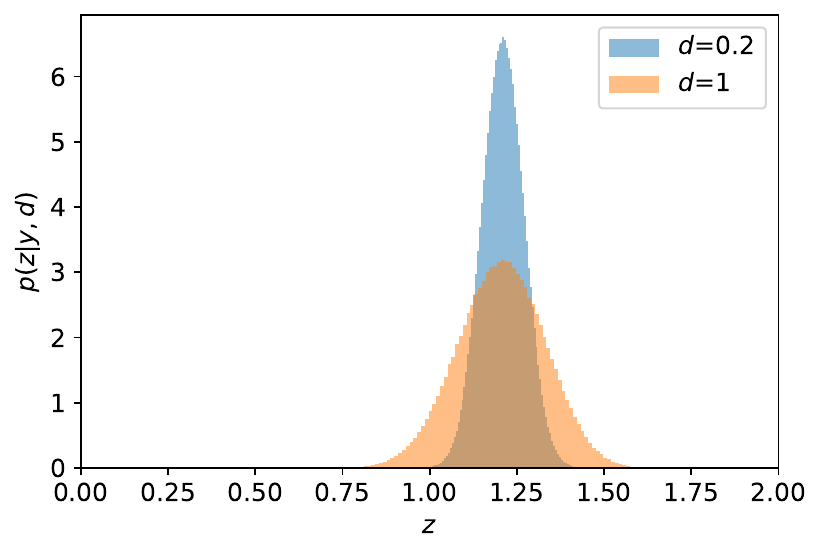}
    \label{fig:posterior_predictives_Param1}
    }%
    \hfill
    \subfloat[Posterior-predictive distributions conditioned on $y$ simulated at $\theta=0.9$
    ]{
    \includegraphics[width=0.46\textwidth]{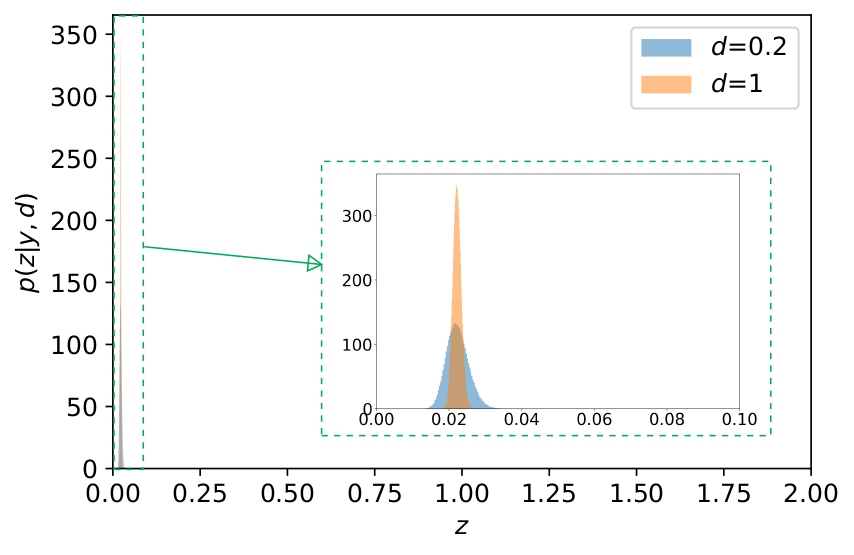}
    \label{fig:posterior_predictives_Param2}
    }%
    \caption{Case T3: Prior-predictive distribution and example posterior-predictive distributions. The bottom-left plot conditions on $y$ simulated at $\theta=0.1$, and $d=0.2$ yields a narrower posterior; the bottom-right plot conditions on $y$ simulated at  $\theta=0.9$, and $d=1.0$ yields a narrower posterior.
    The variability in information gain thus can alter the ranking of the two designs. 
    When repeated for many samples of $\theta$, $y$, and $z$ and taking the expectation, it is the \emph{expected} information gain on $z$ that is higher at $d=0.2$ than at $d=1.0$.}
    \label{fig:posterior_predictives}
\end{figure}

We perform the same assessment for the posterior-predictive distributions, shown in \cref{fig:posterior_predictives}. To begin, we note that the prior-predictive, shown in \cref{fig:prior_predictive},
is no longer uniform after $\theta$ is transformed through the T3 prediction model in \cref{equ:PredT3} but rather highly concentrated towards the two ends and slightly higher on the left side. The subsequent information gains on the QoIs will thus be computed relative to a KDE fit to this prior-predictive distribution. 
In \cref{fig:posterior_predictives_Param1} we see the case where $y$ is simulated at $\theta=0.1$ and design $d=0.2$ yields a narrower posterior-predictive (information gain 2.49) while design $d=1.0$ offers a slightly wider posterior-predictive (information gain 1.76).
\Cref{fig:posterior_predictives_Param2} presents another example for when $y$ is simulated at $\theta=0.9$, where again the opposite is observed (information gain 2.11 at $d=1.0$ is higher than information gain of 1.91 at $d=0.2$).
First, we note that these information gain values are all lower than their $\theta$ information gain counterparts from \cref{fig:sample_posteriors}; this is again consistent with the inequality in \cref{e:z_bijective}. 
Second, the variability in information gain can again alter the ranking of the two designs. 
Upon taking the \emph{expected} information gain over different $\theta$, $y$, and $z$ values leads to an overall higher EIG in $z$ at $d=0.2$ than at $d=1.0$; this ranking is inverted compared to that in the $\theta$ EIG.

\subsection{2D Nonlinear Test Problems}
\label{ss:2D}

\begin{figure}[htbp]
    \centering
    \includegraphics[width=0.6\textwidth]{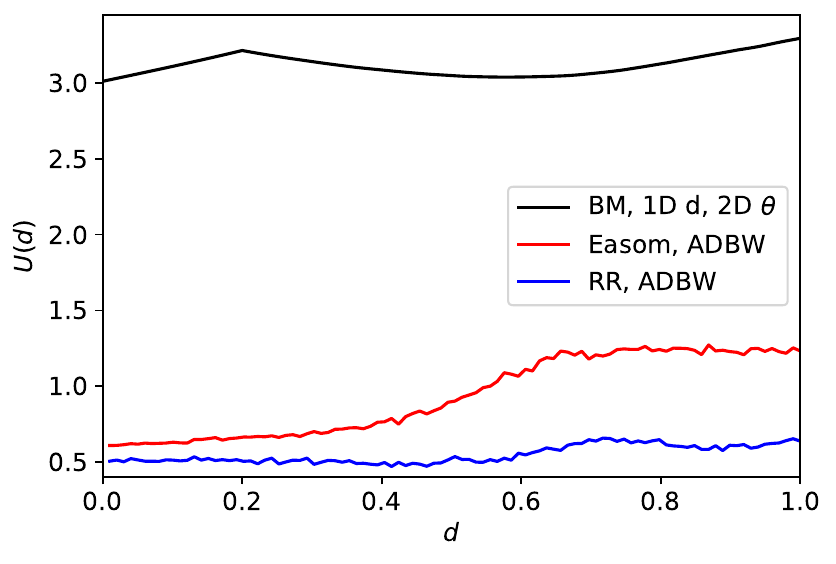}
    \caption{Expected utility comparisons for the 2D $\boldsymbol{\theta}$ case. }
    \label{fig:1Dd2Dt}
\end{figure}

Next, we present test problems that entail 2D parameter, observation, and design spaces. The first example modifies the observation model from \cref{equ:obs1} to incorporate a 2D $\boldsymbol{\theta}$:
\begin{align}
    y = G(\boldsymbol{\theta}, d) + \epsilon = \theta_1^3d^2 + \theta_2 \exp\left( -|0.2-d| \right) + \epsilon
\end{align}
where now the prior becomes $\btheta\sim \CU([0,1]\times[0,1])$ and all other settings remain the same. 
We consider two subcases with new prediction models:
\begin{align}
    z={H}({\btheta}) &= \cos\theta_1  \cos\theta_2 \exp{\left[-(\theta_1-0.4)^2+ (\theta_2-0.6)^2 \right]},  & \text{(Easom Eqn.)} \label{e:Easom}\\ 
    z={H}({\btheta}) &= (1-\theta_1)^2 + 5 (\theta_2-\theta_1^2)^2. & \text{(Rosenbrock Eqn.)}\label{e:Rosenbrock}
\end{align}
The non-GO-OED (via the gridding method) and GO-OED expected utilities are shown in \cref{fig:1Dd2Dt}. 
Both Easom and Rosenbrock subcases maximize $z$ EIG with a design around $d=0.7$.
To understand these results, we see that when $d$ is small $\theta_1$ is not easily identifiable, but as $d$ increases up to $0.2$ both $\theta_1$ and $\theta_2$ improve their identifiablilty, with $\theta_2$'s signal peaking at $d=0.2$. As $d$ further increases, $\theta_1$'s identifiability continues to improve while $\theta_2$'s decreases again. This explains the trend of the non-GO-OED EIG curve in \cref{fig:1Dd2Dt}. 
However, the $d=0.2$ peak disappears for the $z$ EIG curves under the Easom and Rosenbrock prediction models. This can be understood from the forms of \cref{e:Easom,e:Rosenbrock}. As noted above, for small $d$ information is only gained about $\theta_2$ and not $\theta_1$. However, knowing $\theta_2$ alone does not provide much information about these QoIs without knowing some information about $\theta_1$. When $d$ is larger we find more balanced information gain about both parameters alowing more information to be gained about these two functions.

\begin{figure}[htbp]
    \centering
    \subfloat[Easom]{
    \includegraphics[width=0.47\textwidth]{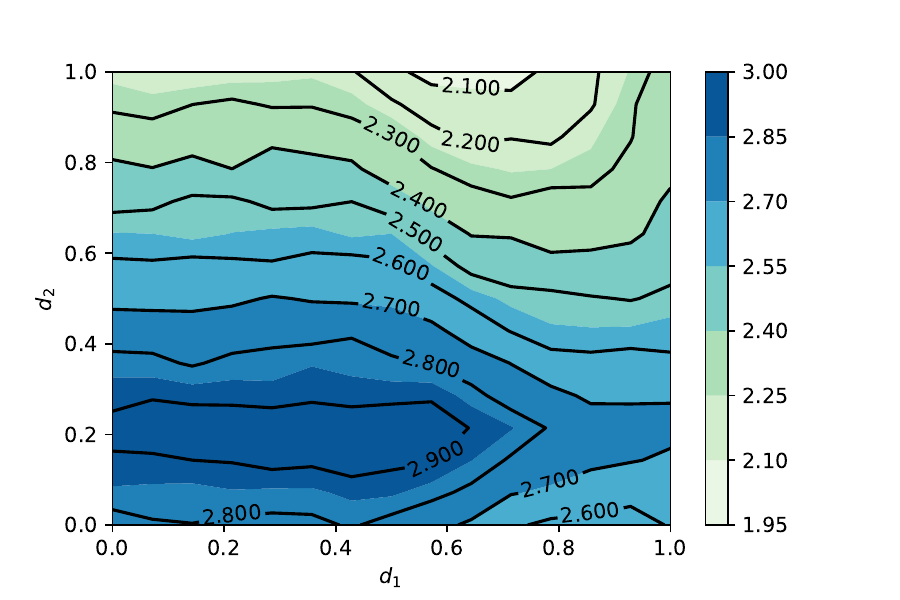}
    }%
    \subfloat[Rosenbrock]{
    \includegraphics[width=0.47\textwidth]{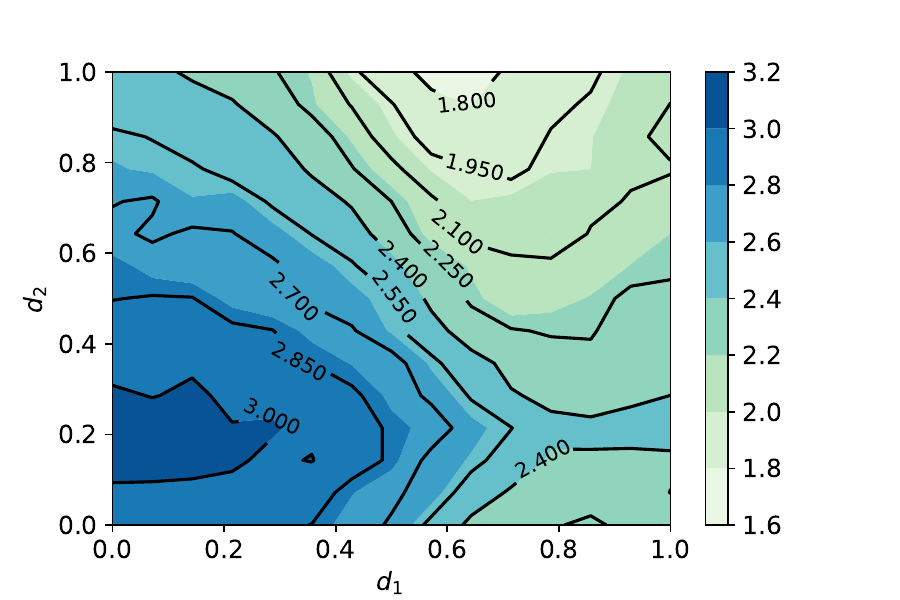}
    }%
    \caption{Expected utility contours for the 2D $\boldsymbol{\theta}$, $\bd$, and $\by$ case.}
    \label{fig:2Dd2Dt}
\end{figure}

Expanding further to 2D design and 2D observation spaces, the next case involves a multi-dimensional observation model:
\begin{align}
\by={\bG}(\btheta,\bd)+\bepsilon= \begin{bmatrix}\theta_1^3 d_1^2 + \theta_2  \exp\left(-\abs{0.2-d_2}\right) \\
        \theta_2^3 d_1^2 + \theta_1  \exp\left(-\abs{0.2-d_2}\right) \end{bmatrix}+\bepsilon
        \label{e:2D_observation}
\end{align}
where $\bepsilon\sim\CN(0,10^{-4}\mathbb{I})$. The prediction models are the same Easom and Rosenbrock equations from \cref{e:Easom} and \cref{e:Rosenbrock}, respectively. The 2D GO-OED expected utility contours are shown in \cref{fig:2Dd2Dt}. 
In contrast to the 1D results in \cref{fig:1Dd2Dt}, the GO-OED contours now exhibit the local maximum at design values around 0.2. This is because the 2D observation model in \cref{e:2D_observation}, through having both $\theta_1$ and $\theta_2$ pre-multiplying the two exponential terms, allows the learning of both parameters well at $d_2=0.2$.  As long as $d_1$ is small, the observation model means that both parameters are well identified so the QoIs will also be well identified.

\begin{figure}[htbp]
    \centering
    \subfloat[BO visited locations]{
    \includegraphics[width=0.47\textwidth]{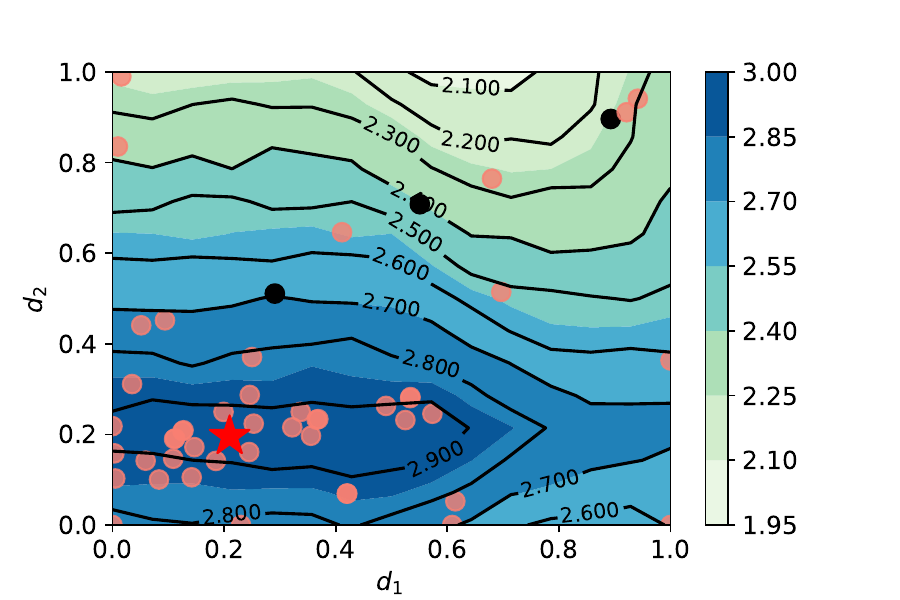}
    \label{fig:BO1}
    }%
    \subfloat[BO convergence history]{
    \includegraphics[width=0.47\textwidth]{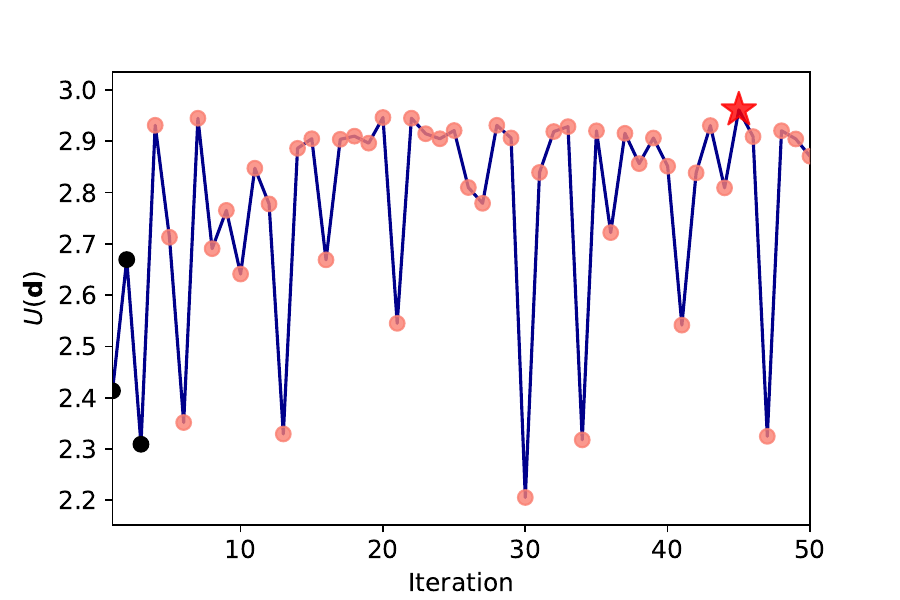}
    \label{fig:BO2}
    }%
    \caption{BO visited locations (left) and optimization convergence history (right) for the Easom prediction model. Black dots are the 3 BO initialization points; orange dots are the visited locations during BO iterations; red star is the numerical optimal design found by BO.}
    \label{fig:OptBO}
\end{figure}

While the expected utility contours have all been constructed so far by estimating $U(\bd)$ on a tensor grid of $\bd$, such brute-force grid search is computationally expensive and may only be done for low dimensional $\bd$. An optimization algorithm, such as the BO algorithm presented in \cref{ss:BO}, would be more efficient to seek out $\mathbf{d}^{\ast}$ directly. 
Here we assess the BO performance in searching for the optimal design via the Easom prediction subcase. 
As seen in \cref{fig:BO1}, BO first randomly select 3 initial points (black) and then uses the UBC acquisition function to select the next points (in orange). We see these points  cluster towards the eventual numerical optimum (red star) but maintains some exploration of the design space as well. The combination of exploitation and exploration can be seen further in \cref{fig:BO2} that plots the convergence history of BO. The $U(\bd)$ values encountered in BO holds an overall increasing trend, but also scattered with low dips that are indicative of the occasional exploration. We will employ BO for all subsequent numerical demonstrations. 

\subsection{Multi-dimensional Nonlinear Test Problem}
\label{ss:ND}

{This example involves $N$-dimensional $\btheta$, $\bd$, $\by$, and one-dimensional $z$, with a focus to study the computational cost as sample sizes and dimensionality vary. The observation model, expressed in component form, is
\begin{align}
y_i = \theta_i^3 d_i^2+ \sum_{j \in \{1,\ldots,N\}, j \neq i}\theta_j \exp{\left(-|0.2-d_j|\right)} + \epsilon_i, \quad i=1,\ldots,N,
\end{align} 
where $\epsilon_i \sim \mathcal{N}(0,10^{-4})$ and prior $\theta_i\sim \mathcal{U}([0,1])$ are all independent, and $d_i\in[0,1]$.
The prediction model is the Rosenbrock function: 
\begin{align}
z=H(\btheta) = \sum_{i=1}^N(1-\theta_i)^2+\sum_{j \in \{1,\ldots,N\}, j\neq i}5(\theta_j-\theta_i^2)^2.
\end{align}}

\begin{figure}[htbp]
    \centering
    \subfloat[Computational cost]{
    \includegraphics[width=0.47\textwidth]{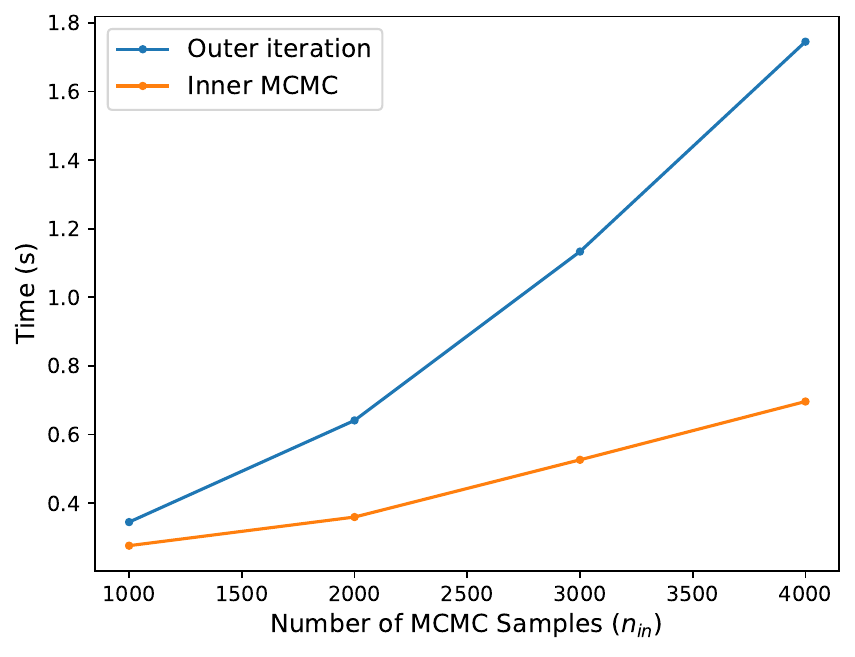}
    \label{fig:Nd_1_cost}
    }%
    \subfloat[MCMC-to-outer-iteration cost ratio]{
    \includegraphics[width=0.47\textwidth]{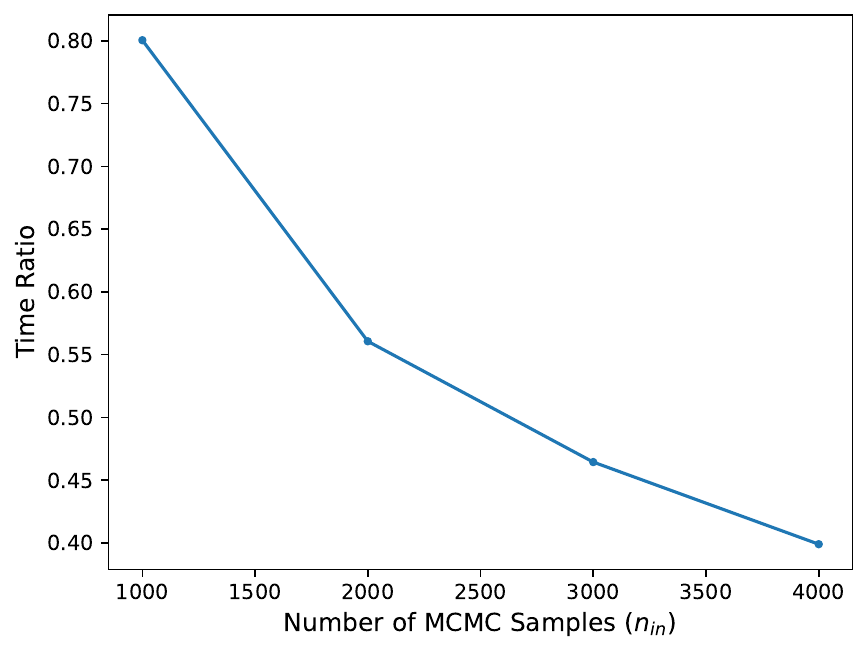}
    \label{fig:ND_1_MCMC_cost}
    }%
    \caption{Computational cost (left) of each outer loop iteration and each inner loop MCMC, and their ratio (right), as functions of $n_{\text{in}}$ while keeping $N=1$ and $n_{\text{out}}=1000$ fixed.}
    \label{fig:ND_1}
\end{figure}

{We begin by analyzing the computational cost as a function of sample sizes, while keeping the dimension fixed at $N=1$. All reported run times are measured on a single core of Intel Xeon Gold Cascade Lake 6248R CPU. Since the total cost scales linearly with the number of outer iteration, we focus on the cost dependence within each outer iteration.
\Cref{fig:Nd_1_cost} presents the per-outer-iteration and per-inner-MCMC run times as functions of $n_{\text{in}}$. Because each outer loop iteration encompasses an MCMC run, the remaining computational time is primarily due to KDE. As expected, MCMC costs increase nearly linearly with $n_{\text{in}}$. However, the cost of outer loop iteration grows at a faster rate. \Cref{fig:ND_1_MCMC_cost} further illustrates this trend by plotting the ratio of MCMC cost to total outer loop iteration cost, which declines sharply as $n_{\text{in}}$ increases. This suggests that the computational cost of KDE scales significantly faster with sample size compared to MCMC, which aligns with our intuition, as KDE requires computing pairwise distances between all samples.
}

\begin{figure}[htbp]
    \centering
    \subfloat[Computational cost]{
    \includegraphics[width=0.47\textwidth]{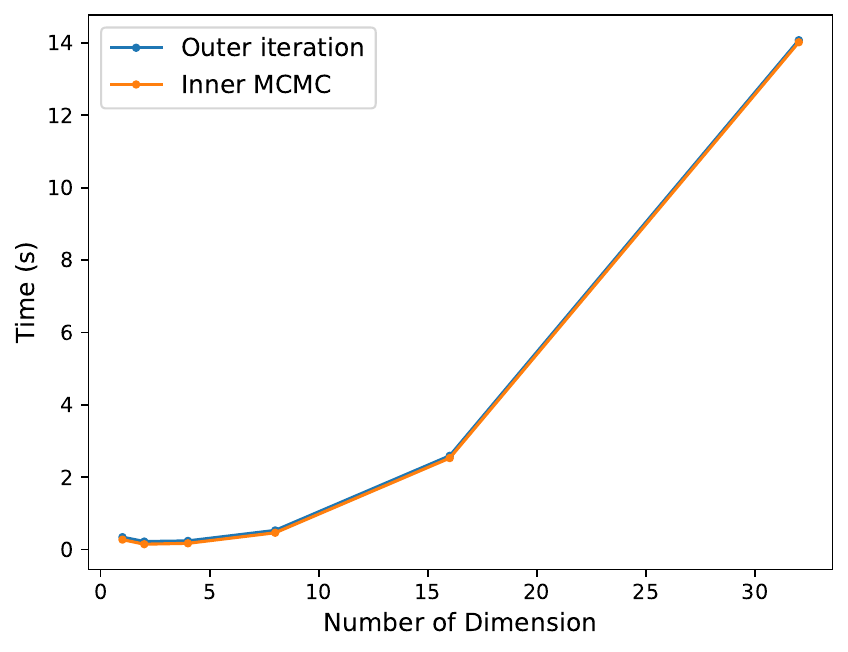}
    \label{fig:ND_cost}
    }%
    \subfloat[MCMC-to-outer-iteration cost ratio]{
    \includegraphics[width=0.47\textwidth]{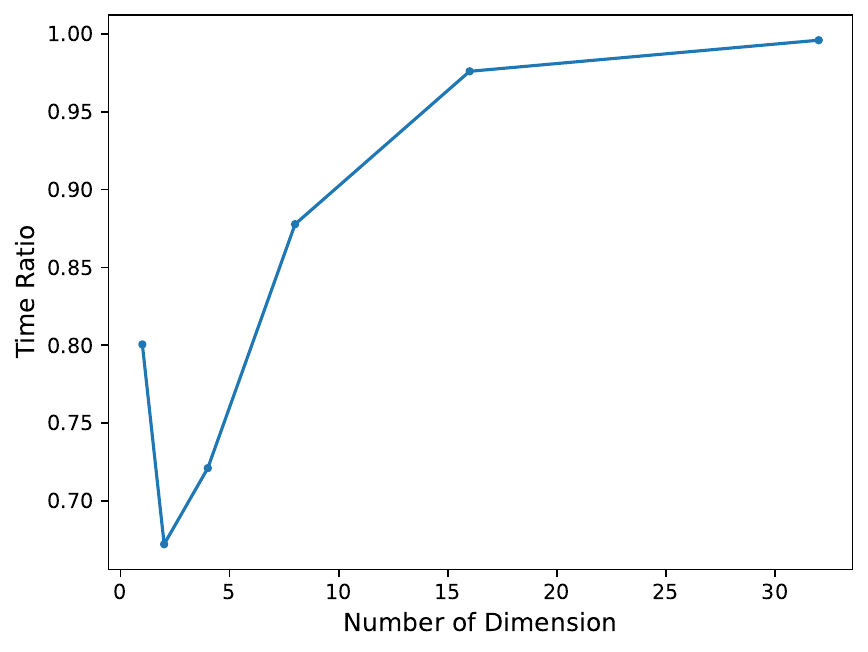}
    \label{fig:ND_MCMC_cost}
    }%
    \caption{Computational cost (left) of each outer loop iteration and each inner loop MCMC, and their ratio (right), as functions of $N$ while keeping $n_{\text{out}}=1000$ and $n_{\text{in}}=1000$ fixed.}
    \label{fig:ND}
\end{figure}

{Next, we study the computational cost as the dimension $N$ varies, while keeping the sample sizes fixed at $(n_{\text{out}},n_{\text{in}})=(1000,1000)$. \Cref{fig:ND_cost} plots the per-outer-iteration and per-inner-MCMC run times, now as functions of $N$. The curves grow super-linearly with $N$. Both curves are very close to each other, suggesting that, under this sample size setting, MCMC computations dominate the overall cost regardless of dimensionality. \Cref{fig:ND_MCMC_cost} further confirms this trend, showing that the ratio of MCMC cost to total outer loop iteration cost to be all greater than 65\% and increasing towards 100\% as $N$ increases. This suggests that MCMC operations are more sensitive to dimensionality than KDE.  }

\subsection{Convection-Diffusion Example}
\label{ss:CD}

In this example, we apply GO-OED for the design of sensor locations in a 2D convection-diffusion field. In this scenario, the concentration $c$ (e.g., of a chemical contaminant) at location $\bx=({x_1,x_2})$ and time $t$ is governed by the convection-diffusion partial differential equation (PDE):
\begin{align}
    \frac{\partial c(\bx,t;\btheta)}{\partial t} = \nabla^2 c - \bu(t) \cdot \nabla c + S(\bx,t;\btheta), \quad \bx \in [-1, 2]^2, \quad t>0
\end{align}
where $\btheta=(\theta_{{x_1}},\theta_{{x_2}})\in [0, 1]^2$ is the (unknown) source location with a uniform prior $\CU([0,1]^2)$, $\bu=(50t,50t)$ is the (known) convection velocity, and the source function $S$ has the form:
\begin{align}
    S(\bx,t;\btheta) = \frac{s}{2\pi h^2}\exp\left(-\frac{\Vert\btheta-\bx\Vert_{2}^2}{2h^2}\right)
\end{align}
with $s=2$ and $h=0.05$ representing source strength and source width respectively. No-flux (homogeneous) Neumann boundary conditions are applied on all four boundaries of the square domain, and the initial condition is $c(\bx,0;\btheta)=0$.

\begin{figure}[htbp]
  \centering
  \includegraphics[width=1.0\textwidth]{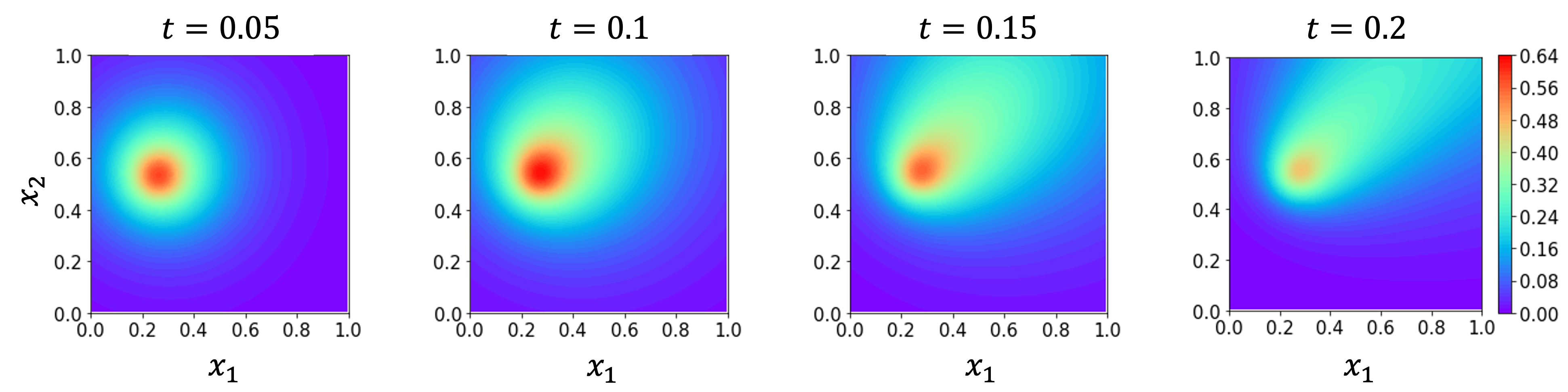}
  \caption{Example numerical solutions of the concentration field at different time snapshots with $\btheta=[0.257,0.528]$. The solution is solved using finite volume in the wider computational domain $[-1,2]^2$ but displayed here in the region of interest $[0,1]^2$.
  }
  \label{fig:Source_exp}
\end{figure}

\paragraph{Numerical Solver and Surrogate Model}
To numerically solve the PDE, we use second-order finite volume method on a uniform grid with $\Delta {x_1} = \Delta {x_2} = 0.01$, and the fractional step method for time marching with stepsize $\Delta t = 5\times 10^{-4}$. The fractional step method combines an explicit second-order Adams--Bashforth discretization for the convective term and an implicit second-order Crank--Nicolson discretization for the diffusive term. Moreover, we employ the QUICK scheme~\cite{Leonard1979} on the convective term to increase numerical stability. Example solutions of the concentration field at different time snapshots are shown in \cref{fig:Source_exp}.

\begin{figure}[htbp]
  \centering
  \subfloat[$t=0.05$]{\label{fig:Source_surrogate_1}\includegraphics[width=0.47\textwidth]{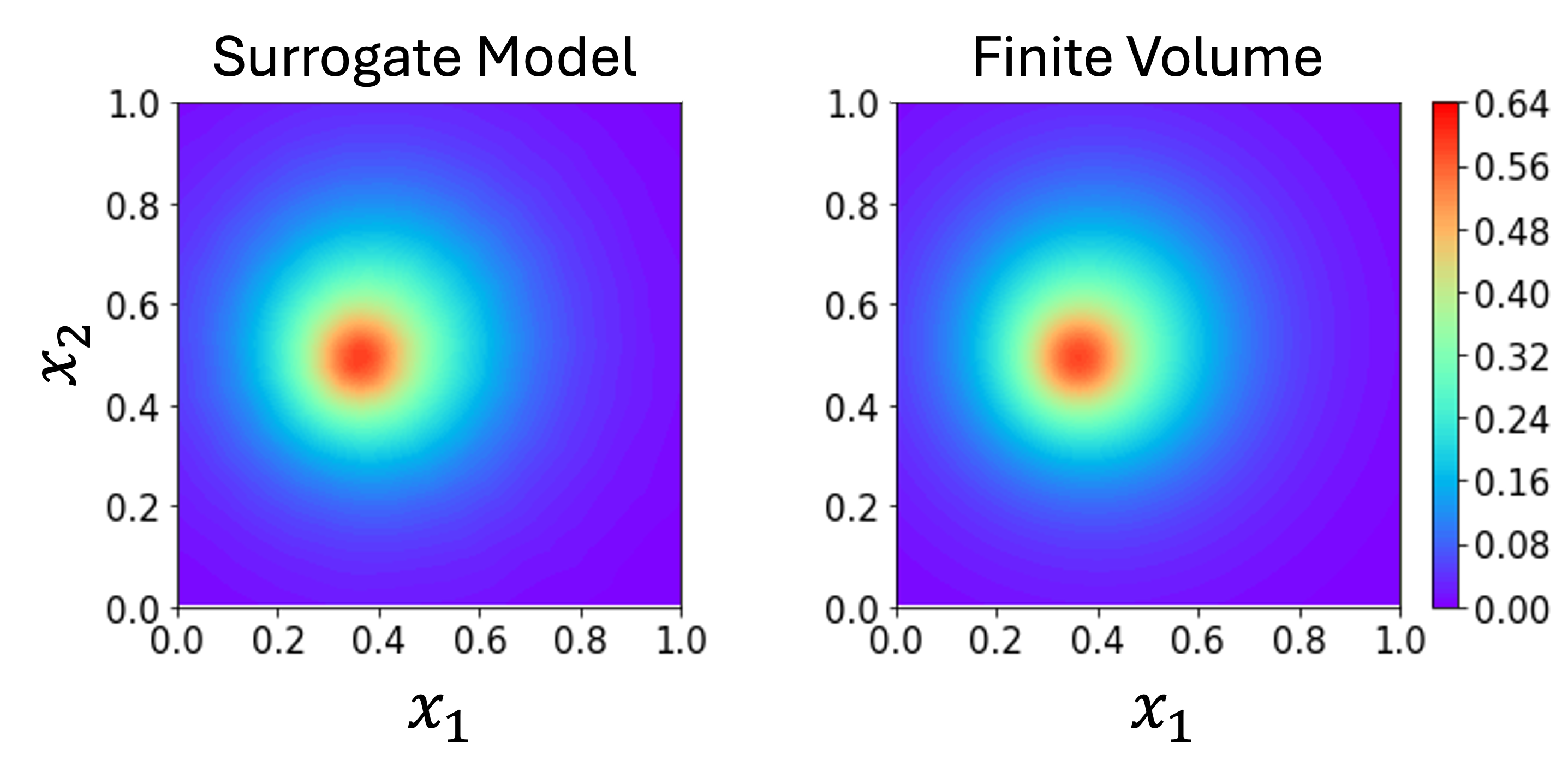}}\hspace{1em}
  \subfloat[$t=0.2$]{\label{fig:Source_surrogate_2}\includegraphics[width=0.47\textwidth]{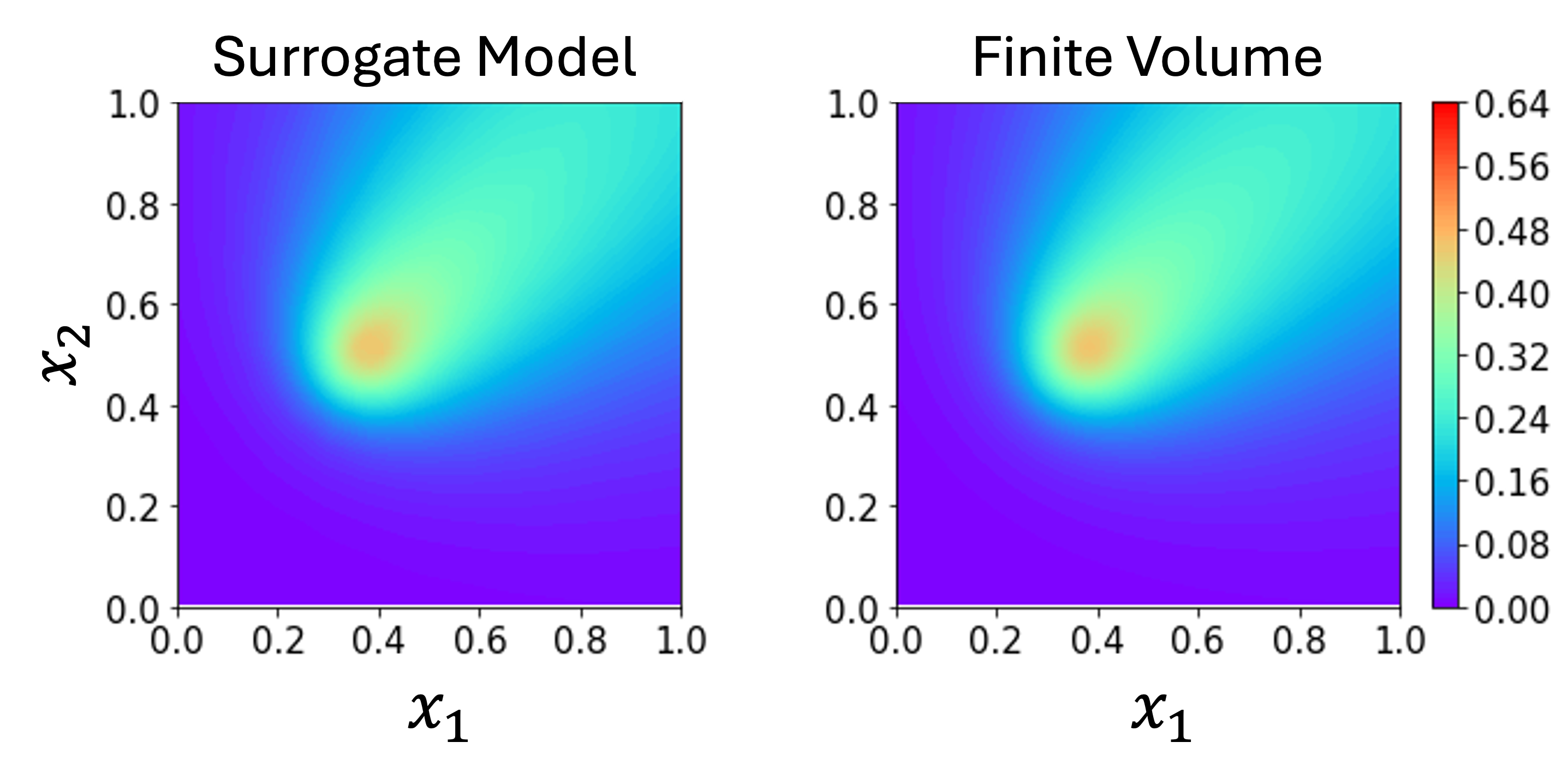}}
  \caption{Example comparisons of the concentration field at $t_1=0.05$ and $t_2=0.2$ with $\btheta=[0.257,0.528]$, obtained using the DNN surrogates (left columns) and finite volume (right columns). 
  }
  \label{fig:Source_surrogate}
\end{figure}

The OED cases (detailed setups to be presented later) will entail 
making observations
at $t_1 = 0.05$ for 
predicting various QoIs 
at future time $t_2=0.2$---hence, the concentrations $c(\bx,t_1,\boldsymbol{\theta})$ and $c(\bx,t_2,\boldsymbol{\theta})$ will be needed.
While it is possible to directly use the finite volume forward model through the entire OED procedure, we construct deep neural network (DNN) surrogate models for quantities $c(\bx,t_1;\btheta)$ and $c(\bx,t_2;\btheta)$ in order to accelerate the computations. 
Each DNN has a four-dimensional input layer taking $\bx$ and $\btheta$, five hidden layers with 40, 80, 40, 20 and 10 nodes, and a scalar output $c$. To facilitate training of these surrogate models, finite volume solution fields are obtained at 2000 random $\btheta$ samples drawn from the prior. The full dataset for the region of interest $\mathbf{x}\in [0,1]^2$ thus entails $2000 \times \frac{(1-0)}{\Delta {x_1}} \times \frac{(1-0)}{\Delta {x_2}} = 2\times10^7$ total points for $c(\bx,t_1,\boldsymbol{\theta})$, and also for $c(\bx,t_2,\boldsymbol{\theta})$. Each dataset is randomly shuffled and then divided to training and testing with an 80--20 split. 
\Cref{fig:Source_surrogate} provides comparisons of the the predicted concentration fields at $t_1$ and $t_2$ using DNN surrogates (left columns) and finite volume (right columns), which agree very well with testing mean-squared errors around $10^{-6}$ and $10^{-7}$ for $c(\bx,t_1;\btheta)$ and $c(\bx,t_2;\btheta)$, respectively. More crucially, the DNN surrogates accelerate each forward model evaluation by about $10^5$ times compared to finite volume.

\begin{figure}[htbp]
    \centering
    \includegraphics[width=0.6\textwidth]{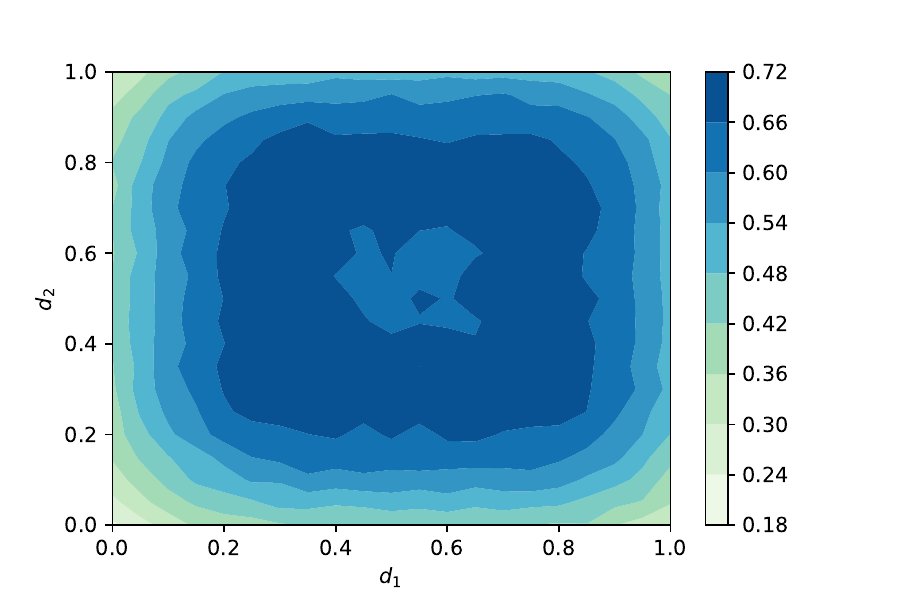}
    \caption{Convection-diffusion 1-sensor design: non-GO-OED expected utility contour.}
    \label{fig:nonGOOED}
\end{figure}

We begin by setting up an observation model for a design problem of selecting the coordinates of a single sensor $\bd \in [0,1]^2$ at which the concentration can be observed:
\begin{align}
    y = G(\boldsymbol{\theta},\bd) + \epsilon =c(\bx=\bd,t_1,\boldsymbol{\theta}) + \epsilon,
\end{align}
with $\epsilon\sim\CN(0,0.05^2\II)$. For a non-GO-OED that seeks to maximize the EIG in $\boldsymbol{\theta}$, the expected utility contour is shown in \cref{fig:nonGOOED} where the optimal design appears around an inner ring roughly radius 0.2 from the center and slightly shifted to the top-right due to the convection direction to the top-right.

\begin{figure}[htbp]
    \centering
    \subfloat[$\bxi=(0.1,1.0)$ (top-left)]{
    \includegraphics[width=0.47\textwidth]{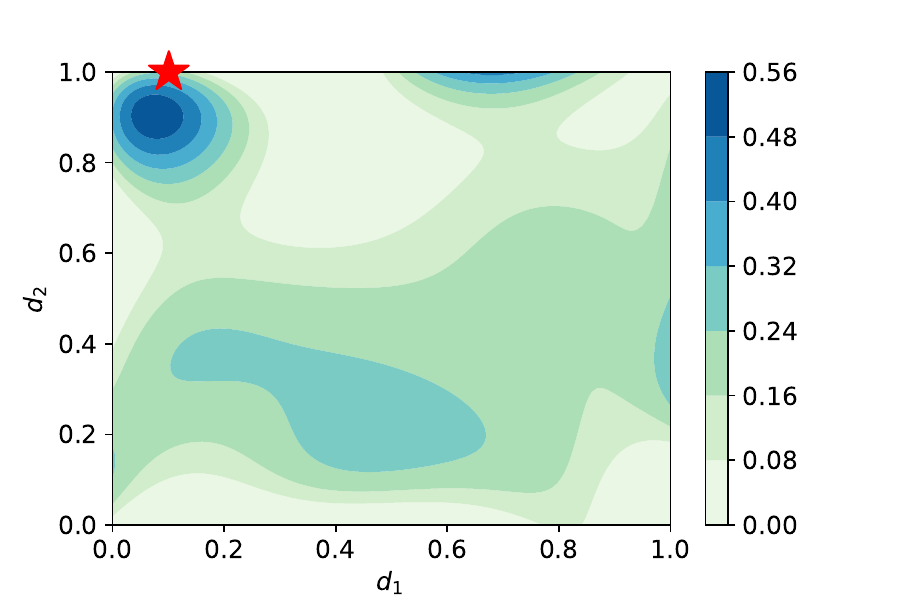}
    \label{fig:CD1DTL}
    }%
    \subfloat[$\bxi=(1.0,1.0)$ (top-right)]{
    \includegraphics[width=0.47\textwidth]{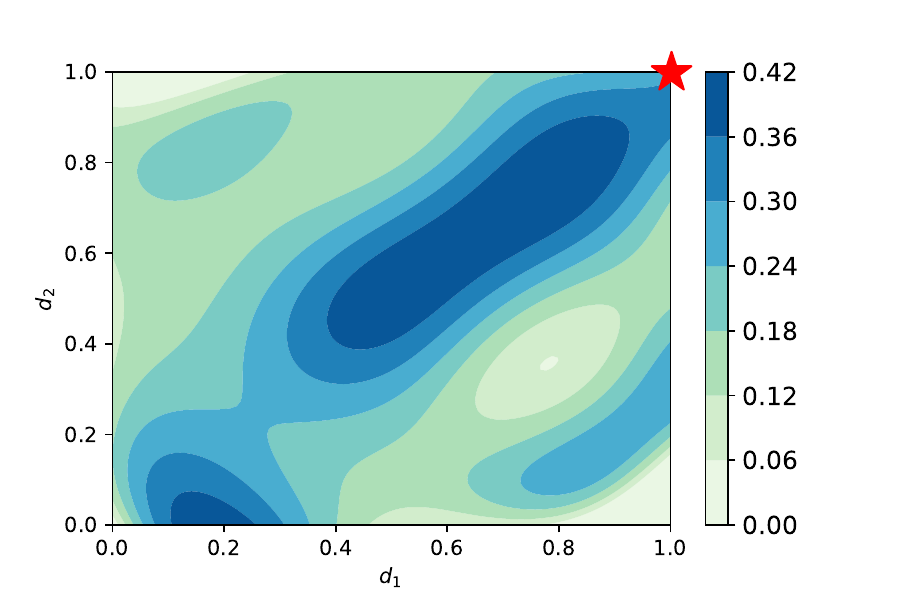}
    \label{fig:CD1DTR}
    }%
    \\
    \subfloat[$\bxi=(0.1,0.1)$ (bottom-left)]{
    \includegraphics[width=0.47\textwidth]{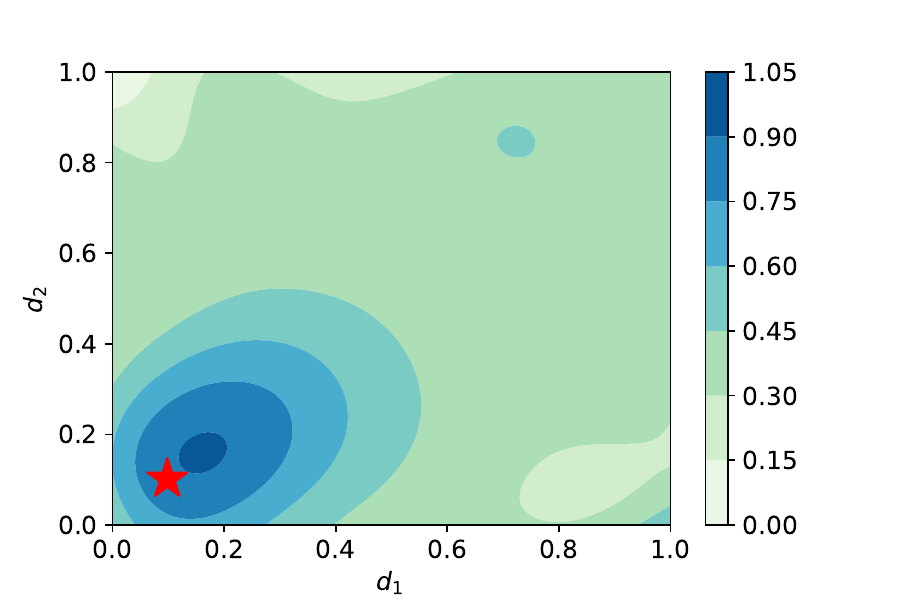}
    \label{fig:CD1DBL}
    }%
    \subfloat[$\bxi=(1.0,0.1)$ (bottom-right)]{
    \includegraphics[width=0.47\textwidth]{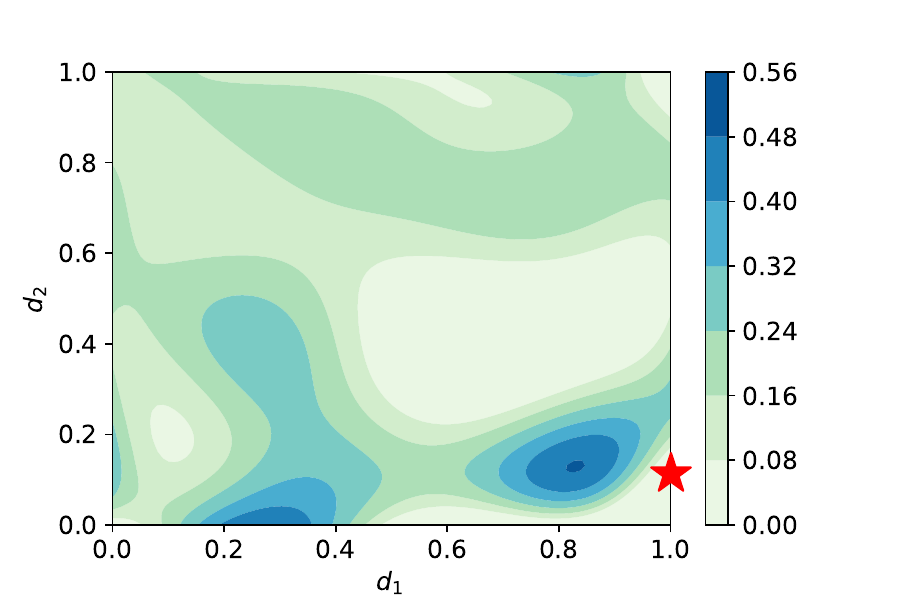}
    \label{fig:CD1DBR}
    }%
    \centering
    \caption{Convection-diffusion 1-sensor design: GO-OED expected utility contours for subcases (a)--(d), where each predictive QoI is the concentration at future time $t_2$ and location $\bxi$ (marked by red star). 
    }
    \label{fig:CD1D}
\end{figure}

\paragraph{Future Concentration QoIs} For GO-OED, we first investigate when the predictive QoI is set to
\begin{align}
    z=H(\boldsymbol{\theta})=c(\bx=\bxi,t_2,\boldsymbol{\theta})
\end{align}
with four subcases of (a) $\bxi=(0.1,1.0)$, (b) $\bxi=(1.0,1.0)$, (c) $\bxi=(0.1,0.1)$, and (d) $\bxi=(1.0,0.1)$. Each subcase corresponds to predicting the concentration near one of the four corners at future time $t_2$. 
The expected utility contours for the subcases are shown in \cref{fig:CD1D}, which appear drastically different from the non-GO-OED result in \cref{fig:nonGOOED}. 
The regions of high expected utility roughly coincide with location of the predictive QoIs: for example, for subcase (a) where the QoI is in near the top-left corner, the GO-OED optimal design also follows towards the top-left. The more elongated contour for subcase (b) results from the convection in the top-right direction, where there is value in taking the earlier measurement at $t_1$ upstream of the targeted top-right $\bxi$ at $t_2$.

\begin{figure}[htbp]
    \centering
    \subfloat[$\bxi_1=(0.08,0.98)$, $\bxi_2=(0.12,0.98)$]{
    \includegraphics[width=0.47\textwidth]{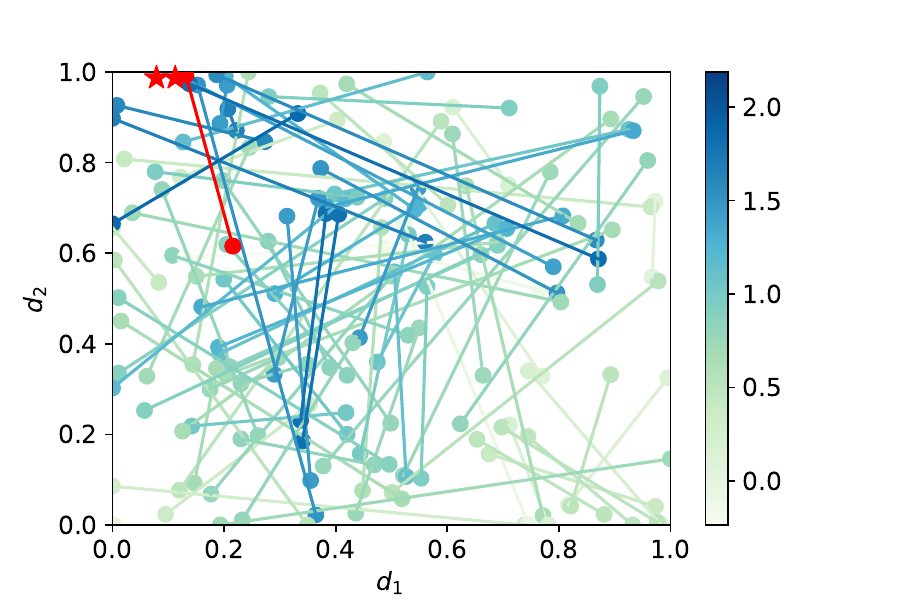}
    \label{fig:CASE_2Pre_1}
    }%
    \subfloat[$\bxi_1=(0.10,1.00)$, $\bxi_2=(1.00,1.00)$]{
    \includegraphics[width=0.47\textwidth]{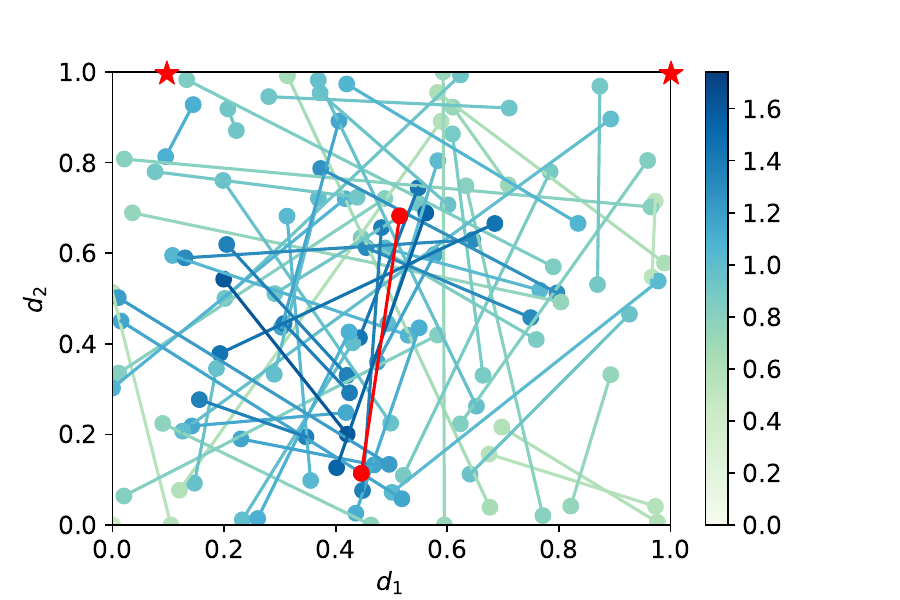}
    \label{fig:CASE_2Pre_2}
    }%
    \caption{Convection-diffusion 2-sensor design: GO-OED for subcases (a) and (b), where the predictive QoIs are the concentration at $t_2$ and two locations $\bxi_1$ and $\bxi_2$ (marked by red stars). The plot shows the sensor combinations encountered in BO (each combination is connected by a straight line) with the optimal combination shown in red; the expected utility value is indicated by color intensity. }
    \label{fig:CD2D}
\end{figure}

\begin{figure}[htbp]
    \centering
    \subfloat[$\bxi_1=(0.08,0.98)$, $\bxi_2=(0.12,0.98)$]{
    \includegraphics[width=0.47\textwidth]{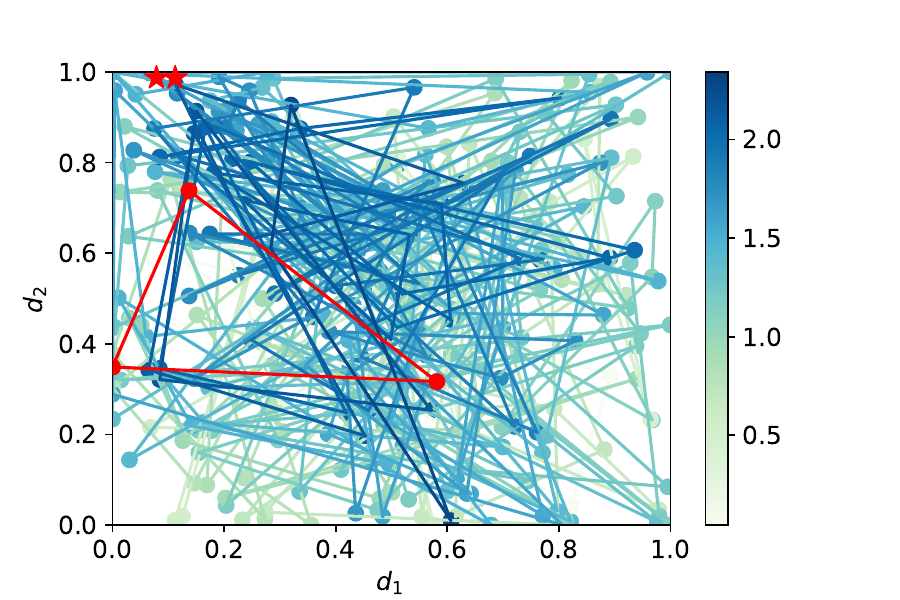}
    \label{fig:CASE_2Pre_3}
    }%
    \subfloat[$\bxi_1=(0.10,1.00)$, $\bxi_2=(1.00,1.00)$]{
    \includegraphics[width=0.47\textwidth]{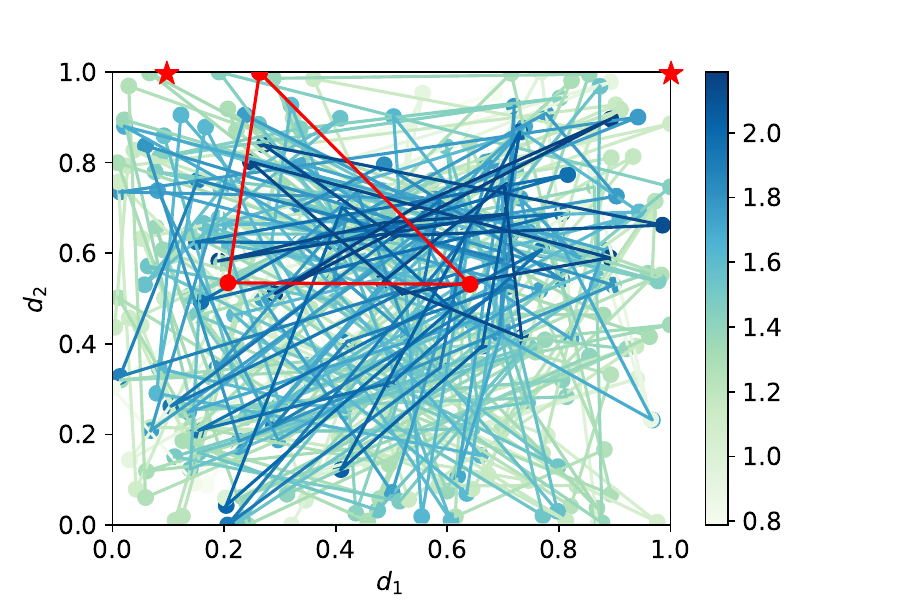}
    \label{fig:CASE_2Pre_4}
    }%
    \caption{Convection-diffusion 3-sensor design: GO-OED for subcases (a) and (b), where the predictive QoIs are the concentration at $t_2$ and two locations $\bxi_1$ and $\bxi_2$ (marked by red stars). The plot shows the sensor combinations encountered in BO (each combination is connected by a triangle) with the optimal combination shown in red; the expected utility value is indicated by color intensity. }
    \label{fig:CD3D}
\end{figure}

Next, we illustrate a few cases with multiple sensors and combinations of QoIs. Consider the GO-OED case where the concentration at two locations, $\bxi_1$ and $\bxi_2$, are of interest:
\begin{align}
    \bz=\bH(\boldsymbol{\theta})=\begin{bmatrix}c(\bx=\bxi_1,t_2,\boldsymbol{\theta}) \\ c(\bx=\bxi_2,t_2,\boldsymbol{\theta})\end{bmatrix}.
\end{align}
We consider two subcases of (a) $\bxi_1=(0.08,0.98)$ and $\bxi_2=(0.12,0.98)$ and (b) $\bxi_1=(0.1,1.0)$ and $\bxi_2=(1.0,1.0)$.
\Cref{fig:CD2D} shows the sensor combinations encountered in BO (each combination is connected by a straight line) and the expected utility value is indicated by color intensity. The combination with the highest expected utility (i.e., the optimal 2-sensor design) is highlighted in red. In subcase (a), while both the QoI locations $\bxi_1$ and $\bxi_2$ are near the top-left corner, we see the sensor combinations with high expected utility tend to spread from top-left towards crossing the diagonal line of $d_1=d_2$. Subcase (b) similarly does not simply place the two sensors near the $\bxi$ locations. The high-value design patterns change further when designing for 3 sensors in \cref{fig:CD3D}, where now the optimal 3-sensor combination is connected by a triangle. Overall, the 3-sensor patterns appear more spread out, possibly as a result of having a larger number of observation opportunities. 
These observations suggest the non-trivial effects from the experiment dynamics and sensor coordination toward the experiment goals. 

\begin{figure}[htbp]
    \centering
    \includegraphics[width=0.6\textwidth]{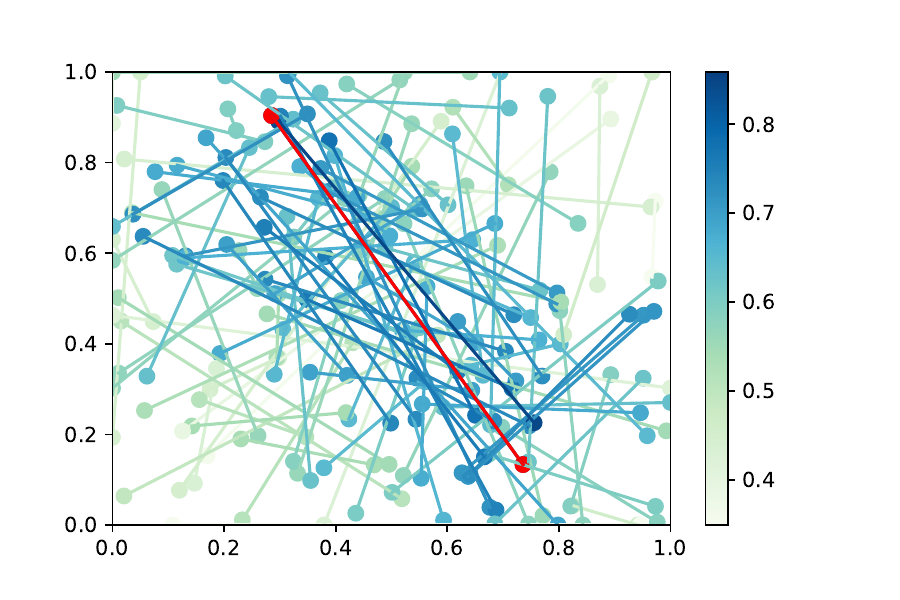}
    \caption{Convection-diffusion 2-sensor design: GO-OED where the predictive QoI is the flux across the right boundary at $t_2$. The plot shows the sensor combinations encountered in BO (each combination is connected by a straight line) with the optimal combination shown in red; the expected utility value is indicated by color intensity.
    }
    \label{fig:CD2DFL}
\end{figure}

\paragraph{Future Flux QoI} Our final cases involve an QoI that is functional of the concentration field, namely the flux through the right boundary of the region of interest at $t_2$. Such quantity is useful for understanding the total contaminant crossing the boundary into a sensitive or protected area to the right. The prediction model becomes
\begin{align}
    z=H(\boldsymbol{\theta})=\int_{-1}^{1} -\left[\frac{\partial c(\bx,t,\boldsymbol{\theta})}{\partial {x_1}}\right]_{(1,{x_2}),t_2,\boldsymbol{\theta}}\,\text{d}{x_2},
\end{align}
which we estimate using second-order center difference upon obtaining $c(\bx,t_2,\boldsymbol{\theta})$ values from the DNN surrogate model. \Cref{fig:CD2DFL} shows the 2-sensor combinations encountered in BO with the best combination highlighted in red. In general, combinations of top-left-to-bottom-right tend to provide higher EIG for this flux QoI, and the sensors are not simply placed geographically close to the flux being considered (i.e., close to the right boundary). 

\begin{figure}[htbp]
    \centering
    \includegraphics[width=0.6\textwidth]{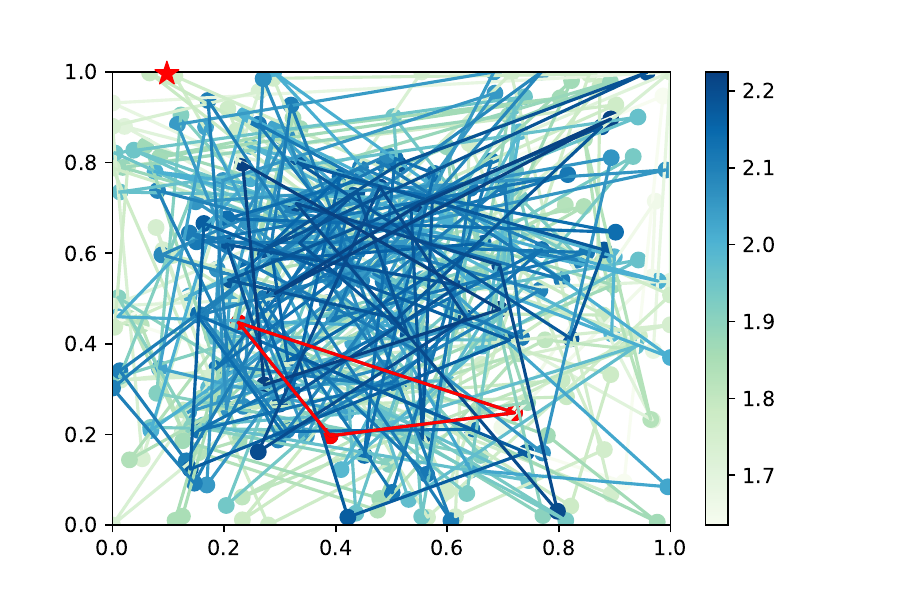}
    \caption{Convection-diffusion 3-sensor design: GO-OED where the predictive QoIs are the concentration at $\bxi=(0.1,1.0)$  and
    flux across the right boundary at $t_2$. The plot shows the sensor combinations encountered in BO (each combination is connected by a triangle) with the optimal combination shown in red; the expected utility value is indicated by color intensity. 
    }
    \label{fig:CD3DFL}
\end{figure}

The last example illustrates a 3-sensor design while combining two different types of QoI together: the concentration at $\bxi=(0.1,1.0)$ and the flux through the right boundary. 
\Cref{fig:CD3DFL} shows the 3-sensor combinations encountered in BO with the best combination highlighted in red. We again observe a more spread-out pattern, likely due of having a larger number of observation opportunities.

\section{Conclusions}
\label{s:conclusions}

We presented a computational method for Bayesian GO-OED that estimates and optimizes the EIG on the predictive QoIs under nonlinear observation and prediction models. This was achieve by establishing a nested MC estimator for the QoIs' EIG, which used MCMC for the necessary posterior sampling. Posterior-predictive samples were then generated by propagating the posterior samples through the prediction model, and subsequently posterior-predictive PDF was approximated via KDE, finally allowing the KL divergence to be computed from the QoIs' prior-predictive distribution to the posterior-predictive distribution. The GO-OED design was then found by maximizing the EIG estimate in the design space using BO.

A number of numerical experiments were provided to illuminate different aspects of the GO-OED framework. These included 1D test problems for validating GO-OED results against alternate computing methods, and exploring GO-OED's numerical behavior (e.g., KDE bandwidth) in simple, controlled settings. 2D examples then followed to demonstrate the effectiveness of BO. {Multi-dimensional cases were explored to examine how computational cost varies with sample size and dimensionality.} Finally, a problem of sensor placement in a convection-diffusion field involving physics-based modeling was investigated to illustrate different predictive QoIs that included concentrations at various locations and flux across a boundary, all at a future time. Throughout these examples, we demonstrated that GO-OED and non-GO-OED design configuration may differ significantly.

A key limitations of this paper's GO-OED is its dependence on MCMC and KDE. While MCMC can be accelerated by initializing the walkers from the true sample-generating parameter values, it can still be slow for complex posterior manifolds and high dimensional parameter spaces. This can introduce high estimator variance. KDE in principle converges to the true density as sample size increases, its finite approximation error can still greatly shift the optimal design locations. The adaptive tuning of bandwidth is also not infallible (i.e., still leading to bias), and becomes expensive if re-tuning is required frequently (e.g., for every MC sample). Promising directions of future work thus entail seeking more efficient methods for estimating the EIG, for example through more sophisticated density estimations such as Gaussian mixture models, transport maps, and normalizing flows; deriving and optimizing bounds for EIG; and building density ratio estimators that can also accommodate implicit likelihood situations.

\section*{Acknowledgments}

WS, TC, and XH 
are supported in part
by the U.S. Department of Energy, Office of Science, Office of Office of Advanced Scientific Computing Research under Award Number 
DE-SC0021397. 
Sandia National Laboratories is a multimission laboratory managed and operated by National Technology and Engineering Solutions of Sandia, LLC, a wholly owned subsidiary of Honeywell International, Inc., for the U.S. Department of Energy's National Nuclear Security Administration under contract DE-NA-0003525. This paper describes objective technical results and analysis. Any subjective views or opinions that might be expressed in the paper do not necessarily represent the views of the U.S. Department of Energy or the United States Government.

\bibliographystyle{abbrv}
\bibliography{references}

\begin{thebibliography}{10}

\bibitem{Alexanderian2021}
A.~Alexanderian.
\newblock {Optimal experimental design for infinite-dimensional Bayesian inverse problems governed by PDEs: A review}.
\newblock {\em Inverse Problems}, 37(4):043001, 2021.

\bibitem{Andrieu2003}
C.~Andrieu, N.~de~Freitas, A.~Doucet, and M.~I. Jordan.
\newblock {An Introduction to MCMC for machine learning}.
\newblock {\em Machine Learning}, 50:5--43, 2003.

\bibitem{Atkinson2007}
A.~C. Atkinson, A.~N. Donev, and R.~D. Tobias.
\newblock {\em {Optimum Experimental Designs, With SAS}}.
\newblock Oxford University Press, 2007.

\bibitem{Attia2018}
A.~Attia, A.~Alexanderian, and A.~K. Saibaba.
\newblock {Goal-oriented optimal design of experiments for large-scale Bayesian linear inverse problems}.
\newblock {\em Inverse Problems}, 34(9):aad210, 2018.

\bibitem{Barber2003}
D.~Barber and F.~Agakov.
\newblock {The {IM} algorithm: A variational approach to information maximization}.
\newblock In {\em Advances in Neural Information Processing Systems 16}, pages 201--208. MIT Press, 2003.

\bibitem{Beck2018}
J.~Beck, B.~M. Dia, L.~F. Espath, Q.~Long, and R.~Tempone.
\newblock {Fast Bayesian experimental design: Laplace-based importance sampling for the expected information gain}.
\newblock {\em Computer Methods in Applied Mechanics and Engineering}, 334:523--553, 2018.

\bibitem{Bernardo1979}
J.~M. Bernardo.
\newblock {Expected information as expected utility}.
\newblock {\em The Annals of Statistics}, 7(3):686--690, 1979.

\bibitem{Various2011}
S.~Brooks, A.~Gelman, G.~Jones, and X.-L. Meng, editors.
\newblock {\em {Handbook of Markov Chain Monte Carlo}}.
\newblock Chapman and Hall/CRC, 2011.

\bibitem{Butler2018}
T.~Butler, J.~Jakeman, and T.~Wildey.
\newblock {Combining push-forward measures and Bayes' rule to construct consistent solutions to stochastic inverse problems}.
\newblock {\em SIAM Journal on Scientific Computing}, 40(2):A984--A1011, 2018.

\bibitem{Butler2018a}
T.~Butler, J.~Jakeman, and T.~Wildey.
\newblock {Convergence of probability densities using approximate models for forward and inverse problems in uncertainty quantification}.
\newblock {\em SIAM Journal on Scientific Computing}, 40(5):A3523--A3548, 2018.

\bibitem{Butler2020}
T.~Butler, J.~D. Jakeman, and T.~Wildey.
\newblock {Optimal experimental design for prediction based on push-forward probability measures}.
\newblock {\em Journal of Computational Physics}, 416:109518, 2020.

\bibitem{cao_comparative_1994}
R.~Cao, A.~Cuevas, and W.~Gonz{\'a}lez~Manteiga.
\newblock A comparative study of several smoothing methods in density estimation.
\newblock {\em Computational Statistics \& Data Analysis}, 17(2):153--176, 1994.

\bibitem{catanach2018bayesian}
T.~A. Catanach and J.~L. Beck.
\newblock {Bayesian updating and uncertainty quantification using sequential tempered MCMC with the rank-one modified Metropolis algorithm}.
\newblock {\em arXiv preprint}, arXiv:1804.08738, 2018.

\bibitem{Chaloner1995}
K.~Chaloner and I.~Verdinelli.
\newblock {Bayesian experimental design: A review}.
\newblock {\em Statistical Science}, 10(3):273--304, 1995.

\bibitem{Chopin2020}
N.~Chopin and O.~Papaspiliopoulos.
\newblock {\em {An Introduction to Sequential Monte Carlo Methods}}.
\newblock Springer Nature Switzerland, 2020.

\bibitem{cowles1996markov}
M.~K. Cowles and B.~P. Carlin.
\newblock {Markov chain Monte Carlo convergence diagnostics: A comparative review}.
\newblock {\em Journal of the American Statistical Association}, 91(434):883--904, 1996.

\bibitem{Duong2023}
D.-L. Duong, T.~Helin, and J.~R. Rojo-Garcia.
\newblock {Stability estimates for the expected utility in Bayesian optimal experimental design}.
\newblock {\em Inverse Problems}, 39(12):125008, 2023.

\bibitem{earl2005parallel}
D.~J. Earl and M.~W. Deem.
\newblock Parallel tempering: Theory, applications, and new perspectives.
\newblock {\em Physical Chemistry Chemical Physics}, 7(23):3910--3916, 2005.

\bibitem{Englezou2022}
Y.~Englezou, T.~W. Waite, and D.~C. Woods.
\newblock Approximate {L}aplace importance sampling for the estimation of expected {S}hannon information gain in high-dimensional {B}ayesian design for nonlinear models.
\newblock {\em Statistics and Computing}, 32(5):82, 2022.

\bibitem{Fedorov1972}
V.~V. Fedorov.
\newblock {\em {Theory of Optimal Experiments}}.
\newblock Academic Press, 1972.

\bibitem{Feng2019}
C.~Feng and Y.~M. Marzouk.
\newblock {A layered multiple importance sampling scheme for focused optimal Bayesian experimental design}.
\newblock {\em arXiv preprint}, arXiv:1903.11187, 2019.

\bibitem{foreman-mackey_emcee_2013}
D.~Foreman-Mackey, D.~W. Hogg, D.~Lang, and J.~Goodman.
\newblock emcee: {The} {MCMC} {hammer}.
\newblock {\em Publications of the Astronomical Society of the Pacific}, 125(925):306--312, 2013.

\bibitem{Foster2019}
A.~Foster, M.~Jankowiak, E.~Bingham, P.~Horsfall, Y.~W. Teh, T.~Rainforth, and N.~Goodman.
\newblock {Variational Bayesian optimal experimental design}.
\newblock In H.~Wallach, H.~Larochelle, A.~Beygelzimer, F.~d\textquotesingle Alch\'{e}-Buc, E.~Fox, and R.~Garnett, editors, {\em Advances in Neural Information Processing Systems 32}, pages 14036--14047. Curran Associates, 2019.

\bibitem{frazier2018tutorial}
P.~I. Frazier.
\newblock Bayesian optimization.
\newblock {\em INFORMS TutORials in Operations Research}, 2018:255--278, 2018.

\bibitem{goodman_ensemble_2010}
J.~Goodman and J.~Weare.
\newblock Ensemble samplers with affine invariance.
\newblock {\em Communications in Applied Mathematics and Computational Science}, 5(1):65--80, 2010.

\bibitem{Gramacy2020}
R.~B. Gramacy.
\newblock {\em {Surrogates}}.
\newblock Chapman and Hall/CRC, 2020.

\bibitem{Huan2024}
X.~Huan, J.~Jagalur, and Y.~Marzouk.
\newblock {Optimal experimental design: Formulations and computations}.
\newblock {\em Acta Numerica}, 33:715--840, 2024.

\bibitem{Huan2013}
X.~Huan and Y.~M. Marzouk.
\newblock {Simulation-based optimal Bayesian experimental design for nonlinear systems}.
\newblock {\em Journal of Computational Physics}, 232(1):288--317, 2013.

\bibitem{Huan2014}
X.~Huan and Y.~M. Marzouk.
\newblock {Gradient-based stochastic optimization methods in Bayesian experimental design}.
\newblock {\em International Journal for Uncertainty Quantification}, 4(6):479--510, 2014.

\bibitem{jones1998efficient}
D.~R. Jones, M.~Schonlau, and W.~J. Welch.
\newblock Efficient global optimization of expensive black-box functions.
\newblock {\em Journal of Global Optimization}, 13:455--492, 1998.

\bibitem{jones_brief_1996}
M.~C. Jones, J.~S. Marron, and S.~J. Sheather.
\newblock A {brief} {survey} of {bandwidth} {selection} for {density} {estimation}.
\newblock {\em Journal of the American Statistical Association}, 91(433):401--407, 1996.

\bibitem{Kleinegesse2020}
S.~Kleinegesse and M.~U. Gutmann.
\newblock {B}ayesian experimental design for implicit models by mutual information neural estimation.
\newblock In H.~Daum\'{e} and A.~Singh, editors, {\em Proceedings of the 37th International Conference on Machine Learning (ICML 2020)}, volume 119 of {\em Proceedings of Machine Learning Research}, pages 5316--5326. PMLR, 2020.

\bibitem{latz2021generalized}
J.~Latz, J.~P. Madrigal-Cianci, F.~Nobile, and R.~Tempone.
\newblock {Generalized parallel tempering on Bayesian inverse problems}.
\newblock {\em Statistics and Computing}, 31(5):67, 2021.

\bibitem{Leonard1979}
B.~Leonard.
\newblock {A stable and accurate convective modelling procedure based on quadratic upstream interpolation}.
\newblock {\em Computer Methods in Applied Mechanics and Engineering}, 19(1):59--98, 1979.

\bibitem{Lindley1956}
D.~V. Lindley.
\newblock {On a measure of the information provided by an experiment}.
\newblock {\em The Annals of Mathematical Statistics}, 27(4):986--1005, 1956.

\bibitem{Long2013}
Q.~Long, M.~Scavino, R.~Tempone, and S.~Wang.
\newblock {Fast estimation of expected information gains for Bayesian experimental designs based on Laplace approximations}.
\newblock {\em Computer Methods in Applied Mechanics and Engineering}, 259:24--39, 2013.

\bibitem{movckus1975bayesian}
J.~Mo{\v{c}}kus.
\newblock {On Bayesian methods for seeking the extremum}.
\newblock In {\em Optimization Techniques IFIP Technical Conference}, pages 400--404, 1975.

\bibitem{Nocedal2006}
J.~Nocedal and S.~J. Wright.
\newblock {\em {Numerical Optimization}}.
\newblock Springer New York, 2006.

\bibitem{fernando2014BO}
F.~Nogueira.
\newblock {Bayesian Optimization}: Open source constrained global optimization tool for {Python}, 2014.

\bibitem{Overstall2018}
A.~M. Overstall, J.~M. McGree, and C.~C. Drovandi.
\newblock An approach for finding fully {B}ayesian optimal designs using normal-based approximations to loss functions.
\newblock {\em Statistics and Computing}, 28:343--358, 2018.

\bibitem{doi:10.1080/01621459.1990.10475307}
B.~U. Park and J.~S. Marron.
\newblock Comparison of data-driven bandwidth selectors.
\newblock {\em Journal of the American Statistical Association}, 85(409):66--72, 1990.

\bibitem{paulin2019error}
D.~Paulin, A.~Jasra, and A.~Thiery.
\newblock {Error bounds for sequential Monte Carlo samplers for multimodal distributions}.
\newblock {\em Bernoulli}, 25(1):310--340, 2019.

\bibitem{scikit-learn}
F.~Pedregosa, G.~Varoquaux, A.~Gramfort, V.~Michel, B.~Thirion, O.~Grisel, M.~Blondel, P.~Prettenhofer, R.~Weiss, V.~Dubourg, J.~Vanderplas, A.~Passos, D.~Cournapeau, M.~Brucher, M.~Perrot, and E.~Duchesnay.
\newblock Scikit-learn: Machine learning in {P}ython.
\newblock {\em Journal of Machine Learning Research}, 12:2825--2830, 2011.

\bibitem{pelikan1999boa}
M.~Pelikan, D.~E. Goldberg, and E.~Cant\'{u}-Paz.
\newblock {BOA: The Bayesian optimization algorithm}.
\newblock In {\em Proceedings of the 1st Annual Conference on Genetic and Evolutionary Computation (GECCO 1999)}, page 525–532, 1999.

\bibitem{pompe2018framework}
E.~Pompe, C.~Holmes, and K.~{\L}atuszy{\'n}ski.
\newblock {A framework for adaptive MCMC targeting multimodal distributions}.
\newblock {\em arXiv preprint}, arXiv:1812.02609, 2018.

\bibitem{Poole2019}
B.~Poole, S.~Ozair, A.~Van Den~Oord, A.~Alemi, and G.~Tucker.
\newblock On variational bounds of mutual information.
\newblock In {\em Proceedings of the 36th International Conference on Machine Learning (ICML 2019)}, volume~97 of {\em Proceedings of Machine Learning Research}, pages 5171--5180. PMLR, 2019.

\bibitem{Rainforth2023}
T.~Rainforth, A.~Foster, D.~R. Ivanova, and F.~B. Smith.
\newblock {Modern Bayesian experimental design}.
\newblock {\em Statistical Science}, 39(1):100--114, 2024.

\bibitem{Rasmussen2006}
C.~E. Rasmussen and C.~K.~I. Williams.
\newblock {\em {Gaussian Processes for Machine Learning}}.
\newblock The MIT Press, 2006.

\bibitem{Robert2004}
C.~P. Robert and G.~Casella.
\newblock {\em {Monte Carlo Statistical Methods}}.
\newblock Springer New York, 2004.

\bibitem{roy2020convergence}
V.~Roy.
\newblock {Convergence diagnostics for Markov chain Monte Carlo}.
\newblock {\em Annual Review of Statistics and Its Application}, 7:387--412, 2020.

\bibitem{ryan_optimal_2016}
C.~M. Ryan, C.~C. Drovandi, and A.~N. Pettitt.
\newblock Optimal {Bayesian} {experimental} {design} for {models} with {intractable} {likelihoods} {using} {indirect} {inference} {applied} to {biological} {process} {models}.
\newblock {\em Bayesian Analysis}, 11(3):857--883, 2016.

\bibitem{Ryan2016}
E.~G. Ryan, C.~C. Drovandi, J.~M. Mcgree, and A.~N. Pettitt.
\newblock {A review of modern computational algorithms for Bayesian optimal design}.
\newblock {\em International Statistical Review}, 84(1):128--154, 2016.

\bibitem{Ryan2003}
K.~J. Ryan.
\newblock Estimating expected information gains for experimental designs with application to the random fatigue-limit model.
\newblock {\em Journal of Computational and Graphical Statistics}, 12(3):585--603, 2003.

\bibitem{shahriari2015taking}
B.~Shahriari, K.~Swersky, Z.~Wang, R.~P. Adams, and N.~de~Freitas.
\newblock {Taking the human out of the loop: A review of Bayesian optimization}.
\newblock {\em Proceedings of the IEEE}, 104(1):148--175, 2016.

\bibitem{Sisson2018}
S.~A. Sisson, Y.~Fan, and M.~Beaumont.
\newblock {Handbook of Approximate Bayesian Computation}.
\newblock {\em Chapman and Hall/CRC}, 2018.

\bibitem{wang2023recent}
X.~Wang, Y.~Jin, S.~Schmitt, and M.~Olhofer.
\newblock {Recent advances in Bayesian optimization}.
\newblock {\em ACM Computing Surveys}, 55(13s):1--36, 2023.

\bibitem{wu2021efficient}
K.~Wu, P.~Chen, and O.~Ghattas.
\newblock {An efficient method for goal-oriented linear Bayesian optimal experimental design: Application to optimal sensor placement}.
\newblock {\em arXiv preprint}, arXiv:2102.06627, 2021.

\bibitem{zhan2020expected}
D.~Zhan and H.~Xing.
\newblock Expected improvement for expensive optimization: A review.
\newblock {\em Journal of Global Optimization}, 78(3):507--544, 2020.

\end{thebibliography}

\end{document}